\documentclass[useAMS,usenatbib]{mn2e}
\usepackage{graphicx}

\newcommand{\apj}{ApJ}

\newcommand{\aj}{AJ}
\newcommand{\mnras}{MNRAS}

\newcommand{\apjs}{ApJS}
\newcommand{\MC}{\multicolumn}

\newcommand{\Te}{T$_{\rm e}$}

\DeclareRobustCommand{\ion}[2]{%
\relax\ifmmode
\ifx\testbx\f
{\mathrm{#1\,\textsc{#2}}}\else
{\mathrm{#1\,\mathsc{#2}}}\fi
\else\textup{#1\,{\mdseries\textsc{#2}}}%
\fi}
\def\etal{{\it et al.\ }}

\title[The Nature of an Apparent Ring Galaxy]
{On the Nature of the Apparent Ring Galaxy SDSS J075234.33+292049.8}
\author[Noah Brosch et al.]{%
Noah Brosch,$^{1}$\thanks{E-mail: noah@wise.tau.ac.il (NB)}
Alexei Y.\ Kniazev,$^{2,3}$
Alexei Moiseev,$^{4}$
\newauthor
Simon A.\ Pustilnik,$^{4, 5}$ \\
\rule{-4pt}{20pt}
$^{1}$The Wise Observatory and the School of Physics and
Astronomy, the Raymond and  Beverly Sackler Faculty of Exact
Sciences, \\ Tel Aviv University, Tel Aviv 69978, Israel \\
$^{2}$South African Astronomical Observatory, PO Box 9, 7935 Observatory, Cape Town,
South Africa\\
$^{3}$Southern African Large Telescope Foundation, PO Box 9, 7935 Observatory, Cape Town,
South Africa\\
$^{4}$Special Astrophysical Observatory, Nizhnij Arkhyz, Karachai-Circassia, 369167,
Russia \\
$^5$Isaac Newton Institute of Chile, SAO branch, Nizhnij Arkhyz, Russia \\
}

\begin{document}

\date{Accepted 2009 September 03. Received 2009 July 30; in original form 2009 June 15}

\pagerange{\pageref{firstpage}--\pageref{lastpage}} \pubyear{2007}

\maketitle

\label{firstpage}


\begin{abstract}
We show that an object classified as a galaxy in on-line data bases and revealed on sky survey images as a distant ring galaxy is a rare case of polar ring galaxy where the ring is only slightly inclined to the equatorial plane of the central body.
Imaging information from the Sloan Digital Sky Survey (SDSS)
indicates that the diameter of the ring is about 36 kpc.
The SDSS data was combined with long-slit spectroscopic
observations and with  Fabry-P\'{e}rot Interferometer H$\beta$ mapping obtained at
the Russian Academy of Sciences 6-m telescope. We
derived the complex morphologies of this presumed ring galaxy from a
combination of SDSS images and from the kinematical behaviour of the
central body and of the ring, and determined the stellar population
compositions of the two components from the SDSS
colours, from the spectroscopy, and from models of evolutionary stellar synthesis. 
The metallicity of the ring material is slightly under-abundant.
The total luminosity and the total mass of the system
are not extreme, but the rather high M/L$\simeq$20 indicates the
presence of large amounts of dark matter.

We propose two alternative explanations of the appearance of this
object. One is a ring formed by two semi-circular and tight spiral arms
at the end of a central bar. The apparent inclination between the
ring and the central body, and a strange kink at the North-East end
of the ring, could be the result of a warp or of precession of the
ring material. The object could, therefore, be explained as an
extreme SBa(R) galaxy. The other possibility is that we observe a Polar Ring Galaxy
where the inner object is an S0 and the ring is significantly more luminous
than the central object. The compound object would then be similar to the
NGC 4650A galaxy, but then it would be a rare object, with a polar component only modestly inclined to the
equatorial plane of the central body.
Arguments for (and against) both explanations are
given and discussed, with the second alternative being more acceptable.

\end{abstract}


\section*{Introduction}

Galaxies with rings have been studied in order to understand the
internal dynamics and evolution of disk galaxies. This type of
object was the topic of an extensive review by Buta \& Combes
(1996) where they explained that, in most cases, such structures
as rings are associated with bars or other  non-axisymmetric disk
disturbances. The rings are very often sites of active star
formation and, in the majority of cases, result from
some resonance phenomena.

Buta \& Combes mention a small fraction of the galaxies with rings
formed by galaxy collisions, mergers, or accretion of
intergalactic matter. These are called ``collisional ring galaxies'' (CRGs) and were
reviewed by Appleton \& Struck-Marcell (1996); the
Cartwheel galaxy is a good example of this type of objects. The CRGs are formed, according
to the authors, by the ``bulls-eye'' collision of one galaxy with
another that is gas-rich. The collision drives a symmetric density
wave into the gas disk, which triggers star formation in a
circular region centered on the impact site. This ring galaxy formation scenario was first
explored by Lynds \& Toomre (1976) and a catalog of such objects was produced by Madore et al.
(2009). The Madore et al. compilation lists for each entry at least two objects very close
together and having approximately the same redshift that they identify as likely the
collider projectile and the colidee target ring.

Another type of galaxy with rings is the ``accretion ring'' (Buta
\& Combes 1996) of which Hoag's object is deemed to be the prototype. The
literature describes arguments about its nature, whether this is a
ring formed in a barred galaxy whose bar subsequently dissolved
(Brosch 1985) or whether the core is not the remnant of a bar but
an elliptical galaxy (Schweizer et al. 1987) and the ring is the
result of accretion. Schweizer et al. described seven other
galaxies as similar objects and defined a class of ``Hoag-type'' objects
that might be quite rare. Wakamatsu (1990) observed one of those
seven galaxies and showed that its central object is oval,
concluding that it might indeed be the remnant of a bar.

Yet another type of ringed galaxy is the ``polar ring galaxy"
(PRG). Formation mechanisms for such objects were investigated in
detail by Bournaud \& Combes (2003). The PRGs are galaxies that
show outer rings composed of stars, dust and gas, with orbits
approximately perpendicular to the plane of the gas-poor host
galaxy. N-body simulations by Bournaud \& Combes showed that a
scenario by which the ring is formed from material accreted by a
lenticular or elliptical galaxy from another galaxy is a more likely
explanation for the observed PRGs than that of a major galaxy merger
(see also Knappen et al. 2004 and Sil'chenko \& Moiseev 2006).
Alternative formation scenarios were proposed, such as accretion
of cold gas from intergalactic space (Macci{\`o} \etal 2006), or
mergers of two disk galaxies with specific orbital configurations
(Bekki 1998).

Observationally, Iodice \etal (2003) investigated the Tully-Fisher
(T-F) relation for PRGs using HI line widths and K or B-band
luminosities, and showed that these objects are located far from
the T-F relation for spiral galaxies. They concluded that the
larger HI rotation velocities observed for the polar rings, in
comparison with the smaller ones expected from the K-band
luminosity, could be explained if the PRGs would have dark matter
haloes flattened along the equatorial planes of the rings.

Combes (2006) proposed that, since the polar rings orbit at a
different angular orientation than the host galaxy in the same halo, PRGs could
be used as probes of the three-dimensional shape of dark matter
(DM) haloes. This, provided the PRG is in equilibrium in the
gravitational potential and that its mass is not so high as to disturb
the halo gravitational potential. The study of PRGs may,
therefore,  be beneficial also to the investigation of the nature of
the DM in galactic haloes. This was demonstrated by the SALT observations
of the PRG AM1934-563 (Brosch et al. 2007) where the DM halo appears to have collapsed
onto the luminous part of the galaxy, not onto the ring.

The widest sample of PRGs is offered by the compilation of
Whitmore \etal (1990, hereafter referred to as the Polar Ring Catalogue=PRC) with more than
100 objects. 
A sub-sample of galaxies from the PRC was studied by Brocca \etal
(1997). They searched for galaxies that were either up to five
ring diameters away or were of similar size to the ring and could
have encountered it within the last Gyr or so. Since they did not
find more companions of these types in comparison with a normal
galaxy sample, they concluded that it is possible the rings are a
very long-lived phenomenon, formed in the distant past, and not a
recent accretion or interaction event.

Ring galaxies are probably relegated to regions of relatively low galaxy density
and are transitory phenomena. Long-lived rings in clusters of galaxies would contradict the
galaxy harassment of gas-containing objects expected to take place
in cluster environments (Moore 1996, 1998). Moreover, the dimming
of the star-forming ring due to the transformation with time of
its gas and dust into stars would also cause the ring to become
undetectable. Precession of oblique rings in the gravitational
potential of the galaxy, and the dissipation due to the diffuse
material in the ring, should have caused long-lived rings to become
exactly perpendicular to the equatorial planes of their host
galaxies or to cease to exist as separate entities.

The question of identifying high-redshift ring galaxies was addressed for the
first time by Reshetnikov (1997), who found two candidates among the Hubble Deep
Field objects. These two objects, 2-809 at z$\simeq$0.64 and 2-906 at z$\simeq$1.2,
which Reshetnikov found similar to the prototype PRG NGC 2685=the Helix galaxy, have only scant
observational information originating from the four-band HDF observations. A similar
study of the Hubble Ultra-Deep Field (Reshetnikov \& Dettmar 2007) revealed HUDF 1619,
a candidate PRG at z$\simeq$1.3.

A comprehensive survey by Lavery et al. (2004), from a visual
inspection of 162 deep HST archival images, yielded 25 objects
with redshifts from 0.35$\leq$z$\leq$1.0 that could possibly be
ring galaxies. Note that these redshifts are, in most cases,
estimates based on the assumption of a
standard-candle magnitude of M$_{\rm V}$=--21.1 mag for these objects. A comparison
of the estimated vs. measured redshifts for the
six objects with spectroscopy in the Lavery et al. list indicates that two
objects differ very significantly from their estimated redshift
while four are in reasonable agreement. Note also that Lavery et
al. called the objects they studied ``collisional ring galaxies''
following Appleton \& Struck-Marcell (1996); since we believe that
this implies a pre-conception about the nature and origin of the galaxies, we
will refrain from using this term in the present paper.

The object discussed here represents a  first attempt to bridge
between the reasonably well-studied but low redshift PRC sample and those
derived from HST observations by Reshetnikov (1997), Lavery et al. (2004), and
Reshetnikov \& Dettmar (2007). Since the most
distant ring galaxies are few and not well-studied, except by
Lavery et al. (1996, 2004), it may be useful to study objects more
distant than those in the PRG but closer than those found on HST
images that may bridge the distance gap to z$\simeq$0.35. Such objects can serve
also as templates to understand the high-z PRGs, once a larger sample will become available.

This paper presents an interpretation of
existing observational data along with new observations of
the object SDSS J075234.33+292049.8 (referred to as RG1) at z=0.060149, identified as a ring galaxy by EO during an investigation of large-separation gravitationally lensed
AGNs (Ofek 2001, 2002). This object is significantly more distant than most ring galaxies
published and studied so far, with the exception of those in Reshetnikov (1997),
Lavery et al. (2004), and Reshetnikov \& Dettmar (2007).
We show that two possible explanations could be invoked for this object:
a barred spiral, with a pair of semi-circular arms forming the apparent ring,
or a classical but rare PRG, as the immediate morphological evidence would argue.

This paper is organized as follows: \S\ref{txt:Obs_and_Red}  gives a
description of all the observations and data reduction, including
information derived from the SDSS, and long-slit spectra and scanning
Fabry-P\'{e}rot spectroscopy obtained with the 6m BTA telescope of the
Russian Academy of Sciences at the Special Astrophysical Observatory (SAO RAS).
We present our observational results in \S\ref{txt:results}, and analyze them
and present our interpretation of the  object in \S\ref{txt:interp}.
The conclusions drawn from this study are summarized in
\S\ref{txt:summ}.

\begin{figure}
{\centering
 \includegraphics[clip=,angle=0,width=8.5cm]{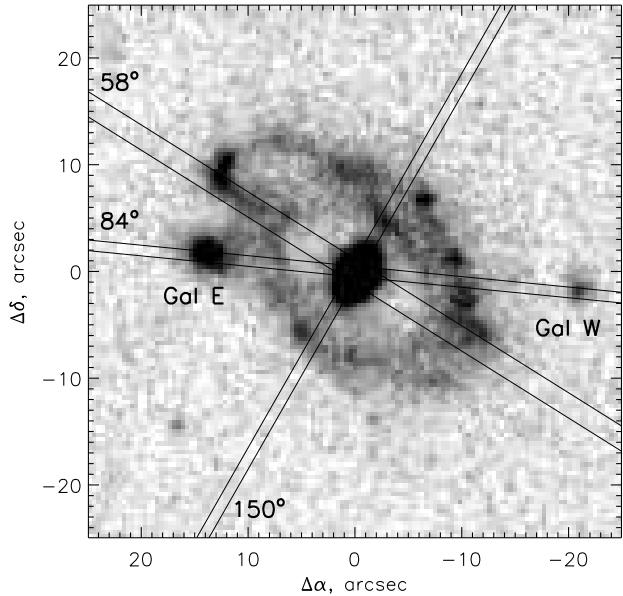}
}
 \caption{%
Deep image of RG1 obtained from a combination of SDSS $g, r$ and $i$ images. North is up and East is
to the left and the image covers approximately 50 arcsec.
The SDSS images were combined with weights \citep{LSB_phot}.
Note the shape of the central object, the apparent doubling of the East side of the ring,
the kink or loop on the top side of the ring, and the impression of a faint extension
along the major axis of the central object that links up to the ring.
Note also the two very nearby galaxies are marked in figure:
E left of the loop at the top of the image and near the ring,
and W on the diametrically opposite side of RG1, more distant and fainter (see also Table~\ref{t:RG1_details}).
Our integral photometry reported in Table~\ref{t:photom} excluded their light contribution.
Slit orientations and actual slit widths for the three position angles used for spectroscopy are also indicated.
Slit position 84$^{\circ}$ explores not only the central object of RG1 and the ring,
but also the two nearby galaxies W and E.
Slit position 58$^{\circ}$ explores the NE kink in the ring and the central object.
Slit position 150$^{\circ}$ explores the minor axis of the ring.
    \label{Fig:RG1_aper}}
\end{figure}

\begin{table}
\caption{Details of the RG1 long-slit observations}
\label{t:Obs}
\begin{tabular}{cccccc} \hline
   Date        & Exposure     & Range          & Slit    &  \MC{1}{c}{PA} & Disp.         \\
	  &  (sec)       &     (\AA)    &('') & ($^{\circ}$)      & (\AA/pix)        \\ \hline
  16.02.2001  & 1200 & 3700--6100     &  2.0    &  57.9           & 2.4          \\
  16.02.2001  & 1200 & 6000--8200     &  2.0    &  57.9           & 2.4          \\
  15.01.2008  & 3600 & 5600--7300     &  1.0    &  84.0           &  0.84 \\
  12.02.2008  & 3600 & 5600--7300     &  1.0    & 150.0           & 0.84 \\ \hline
\end{tabular}
\end{table}

\section{SDSS photometry, BTA observations and data reduction}
\label{txt:Obs_and_Red}

\subsection{SDSS data}
\label{txt:sdss}


The SDSS \citep[][]{York2000,DR5} imaging data were used to provide
morphological parameters for the galaxy, colours of the central
object, and colours and colour distributions for the ring. Since
the source extraction routines of SDSS are not optimized for the
detection and measurement of very faint surface brightness
extended objects, SDSS provides only the photometry for the
central component and also for those ``shreds'' of ring
identified as ``stars'' by the SDSS detection software. Therefore,
specialized photometry in the SDSS bands was applied to measure
the properties of the central component and of the ring
using the low surface brightness detection methods
developed and described in \citet{LSB_phot}.

Specifically, a deep image of the galaxy and of the ring was
created by combining the $g$, $r$, and $i$ images with weights, as
explained in \citet{LSB_phot}. While no physical meaning can easily be
attached to photometry using such a wide and ill-defined
photometric band, the resulting image is significantly deeper than the
individual SDSS images and is useful to investigate the object
morphology. The deep $gri$ combined image is shown in Figure~\ref{Fig:RG1_aper}. 
 An outermost isophote was used to trace
automatically an irregularly-shaped aperture that collects
all the light from even very faint surface brightness parts of
the galaxy. 

At this point, a few remarks about the appearance of RG1 are in order.
Figure~\ref{Fig:RG1_aper} shows a relatively bright, elliptical-shaped central
body encircled by an $\sim$25-arcsec elliptical ring. Two roundish and relatively
compact non-stellar objects are visible near the ring; one is to the East
(galaxy E) and the other is in the opposite side (galaxy W),
slightly more distant from the centre and fainter than galaxy E. These bodies are reminiscent of the proposed ``impactor'' galaxy in a scenario for the  formation of collisional ring galaxies. In addition, the ring shows a loop-like extension or kink to the North-East.

In order to perform integral and surface brightness photometry we
used concentric circular annuli centered on the galaxy's
photo-centre \citep{LSB_phot}, with radii increasing outward by
one SDSS pixel. The average surface brightness and the average colour were determined in
each annulus.

\subsection{Long-slit spectroscopy with the 6m BTA telescope  of the RAS }
\label{txt:long_slit_obs}

First spectroscopic data were obtained 
 with the 6-meter Large Altazimuthal Telescope (BTA)
 of the Special Astrophysical Observatory of Russian
Academy of Science (SAO RAS) on February 16/17 2001 (see Table~\ref{t:Obs} for details).
The 2001 observations used
the Long-Slit spectrograph UAGS (Afanasiev et al. 1995) at the
telescope prime focus equipped with a Photometrics CCD detector
with 1024$\times$1024 pixels each $24\times24\, \mu m^2$ in size.
Two 130\arcsec\ long-slit spectra were obtained in blue and in red
with a grating having 651 grooves/mm yielding a dispersion of
2.4~\AA/pixel and a spectral resolution of 6--7~\AA\,. The slit
position (see Fig.~\ref{Fig:RG1_aper}) was chosen to
cross the central bright region close to the direction of the major
axis of the inclined ring. The wavelength ranges of the obtained
spectra are given in Table~\ref{t:Obs}. A 2\arcsec\ wide slit was
used in both cases. The scale along the slit was 0.39\arcsec/pixel,
very similar to that of the SDSS images.
Reference spectra of an Ar--Ne--He lamp were recorded before or
after each observation to provide  wavelength calibration.
Spectrophotometric standard stars from Bohlin (1996) were observed
for flux calibration.

New spectral observations with the BTA were obtained by AM on January 15/16 and
on February 12/13, 2008, as also detailed in Table~\ref{t:Obs}.
These were collected using the long-slit mode of the SCORPIO
universal focal reducer (Zasov et al.  2008). A volume-phase
holographic grating with 1200 grooves mm$^{-1}$ was used, providing a dispersion
of 0.84 \AA\,pixel$^{-1}$ and yielding spectra that covered
the range $\sim$5600\AA\, to $\sim$7310\AA. The spectral resolution of these spectra was
$\sim$5 \AA\, and the spatial sampling was 0.36 arcsec pixel$^{-1}$. The locations of the
entrance slit for the two observed positions are shown in
Figure~\ref{Fig:RG1_aper}.

The primary data reduction included cosmic-ray
removal with the {\tt MIDAS}\footnote{{\tt MIDAS} is an acronym for the
European Southern Observatory package --- Munich Image Data
Analysis System.} software package, and bias subtraction and
flat-field correction with the {\tt IRAF}\footnote{{\tt IRAF}: the Image
Reduction and Analysis Facility is distributed by the National
Optical Astronomy Observatories, which is operated by the
Association of Universities for Research in Astronomy, In. (AURA)
under cooperative agreement with the National Science Foundation
(NSF).} software package. For subsequent reduction of the
long-slit spectra we used {\tt IRAF}.
In case of UAGS data, after wavelength mapping and night sky subtraction, each 2D
frame was corrected for atmospheric extinction and was flux calibrated.
2D spectra of different spectral regions were combined after that stage.

\begin{figure*}
{\centering
 \includegraphics[clip=,angle=-90,width=17.0cm]{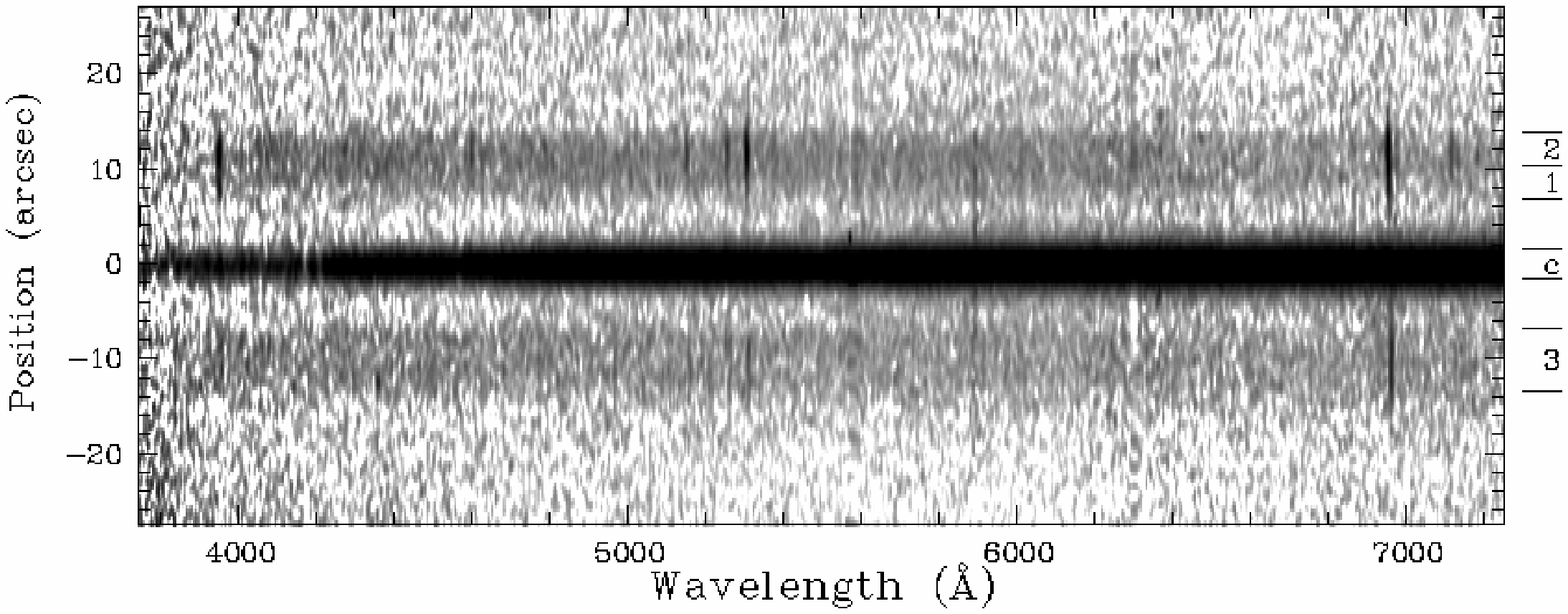}
 \includegraphics[clip=,angle=-90,width=8.0cm]{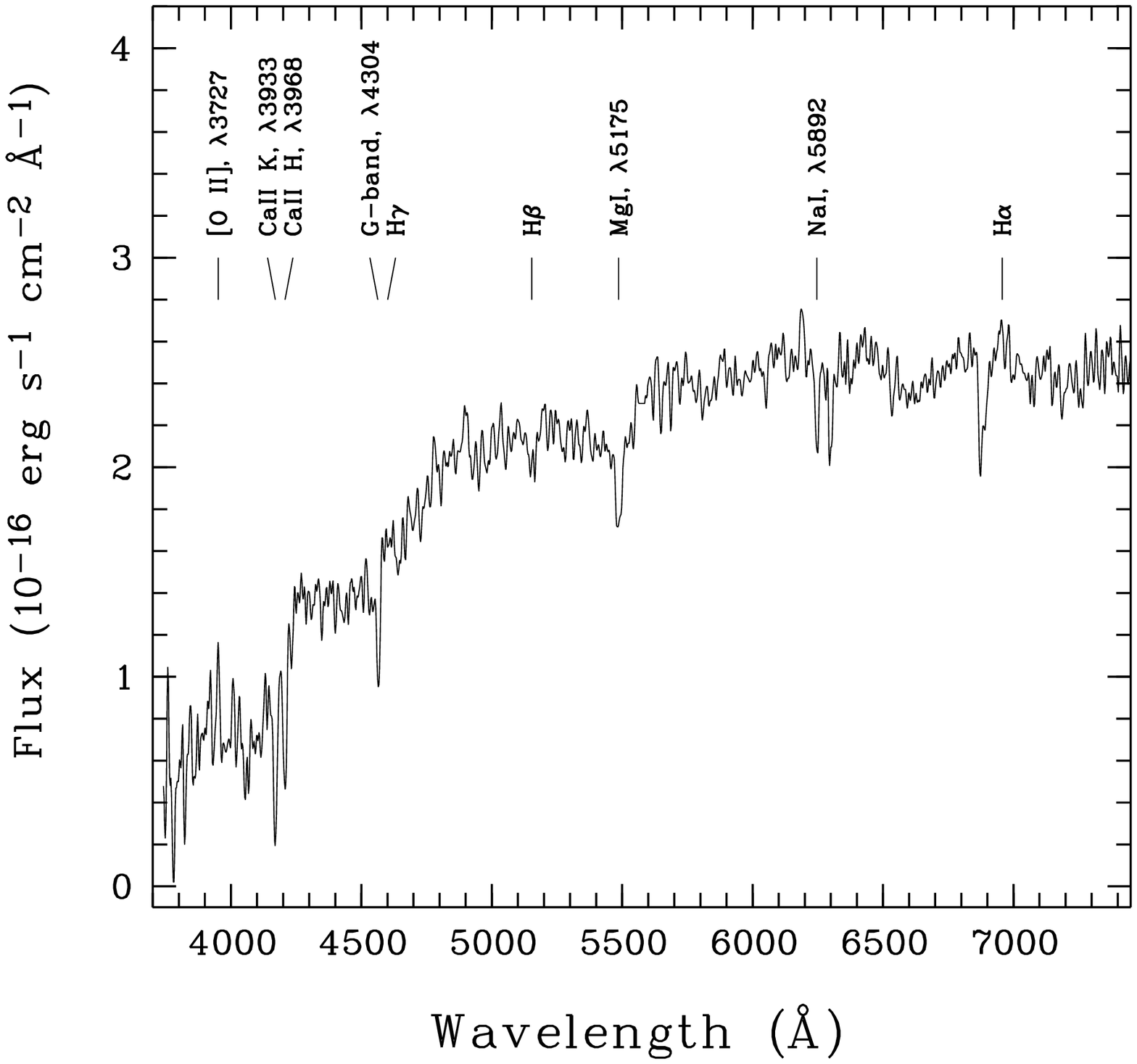}
 \includegraphics[clip=,angle=-90,width=8.0cm]{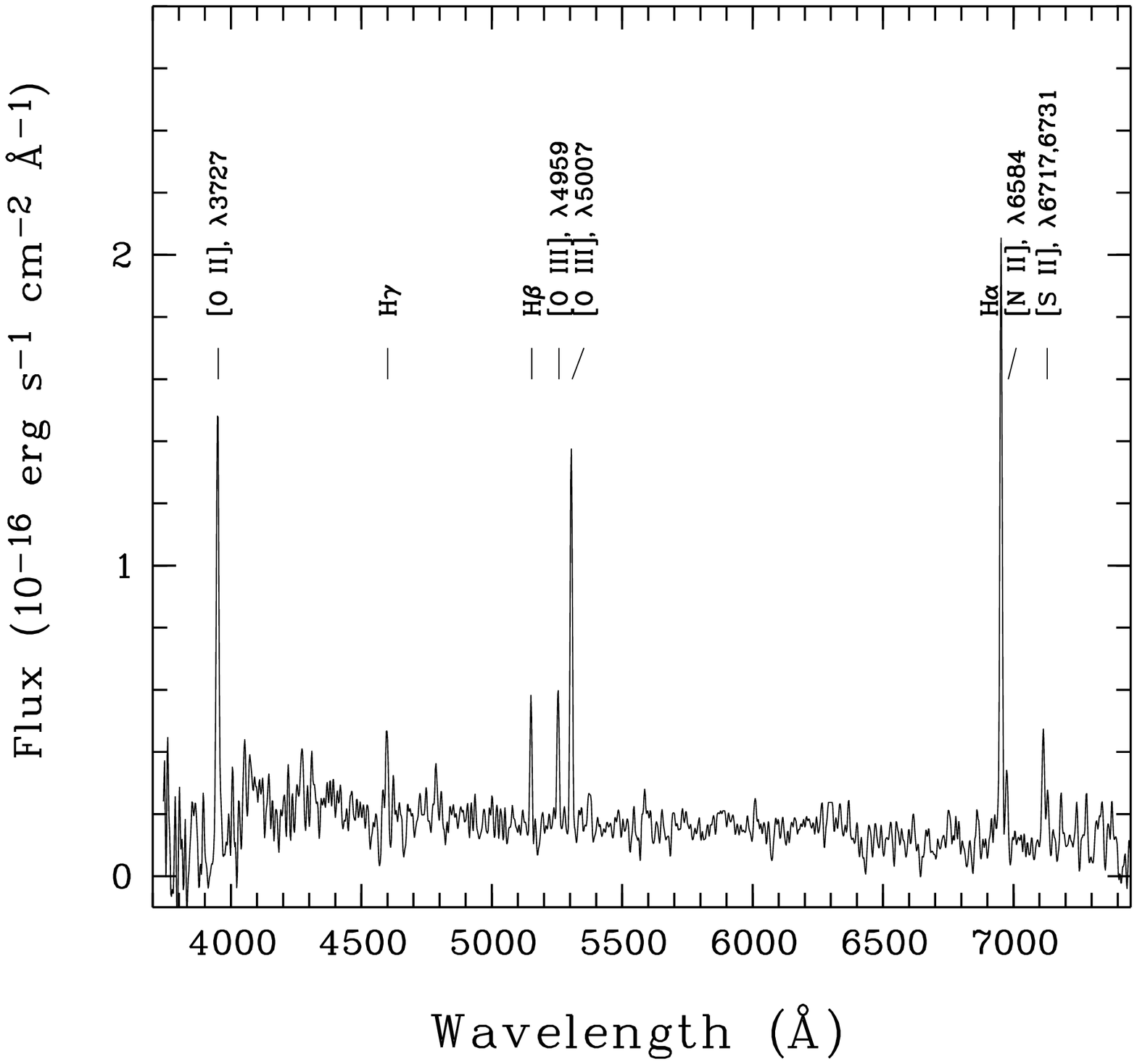}
}
 \caption{{\it Upper panel:}
Part of the reduced  BTA UAGS 2D spectrum of RG1 for $\rm PA \sim 58^{\circ}$.
NE is up. The two arcsec wide slit was positioned on the central
component and along the major axis. The ring region exhibits the
redshifted [\ion{S}{ii}], H$\alpha$, [\ion{N}{ii}],  [\ion{O}{iii}],
H$\beta$, H$\gamma$, and [\ion{O}{ii}] emission lines with measurable
intensities on a visible, though faint, continuum that seems
stronger on the blue side of the image.
The two Ca II H \& K lines and the CH G-band absorption are
visible near the blue end of the central object spectrum. The spectrum of the
central object and of both sides of the ring is easily visible for a
distance of $\pm$14\arcsec\ along the slit.  At the adopted distance of 250 Mpc
1\arcsec\ = 1.2 kpc and the image vertical extent is $\sim$33.6 kpc.
Horizontal lines to the right of the panel show parts of 2D spectrum,
which were averaged and used in Section~\ref{txt:abund}.
{\it Lower panel:} The 1D spectra extracted from the 2D spectrum
observed at $\rm PA \sim 58^{\circ}$:
the central component of RG1 (left, region {\it c}) and
outer part of the ring (right, region {\it 2}).
Some detected emission and absorption lines have been marked. The
spectrum of the central component is equivalent with the one in the SDSS database.
    \label{Fig:2D-spectrum}}
\end{figure*}

RG1 has an SDSS spectrum in the public archives. Since this was
obtained with a single fiber, it samples only the central three arcsec of the
central body. We extracted the innermost three arcsec spectrum from our long-slit
reduced UAGS 2D spectrum and compared it with the SDSS spectrum.
Although the comparison shows that the two spectra match within 10\%, we assumed
that the SDSS spectrum has a more accurate flux distribution and constructed the correction
function based on it.  The reason for this assumption is that SDSS is
observing simultaneously with 640 fibers, where some are targeted on flux standards.
This implies that each SDSS spectroscopic measurement has a few standard stars out of the
640 spectra collected and their fluxes are used for calibration.
For 6m blue and red parts of the final spectrum where observed separately in not-so-perfect weather
conditions. Additionally, blue and red parts of the flux standards were observed separately.
Thus, from our point of view, ``many and simultaneous'' calibrations are better than ``few and separate'' and
we selected the SDSS spectrum to derive the correction function.
The derived function was applied to the reduced UAGS 2D spectrum and the result is
shown in Figure~\ref{Fig:2D-spectrum}
together with the 1D spectra of the central component and the outer part of the ring.


To obtain the line-of-sight velocity distribution
along the slit we used a classical cross-correlation method described in detail by \citet{Zasov00}.
All emission lines were measured with the MIDAS programs
described in detail in \citet{SHOC,Sextans}.

\subsection{Observations with the scanning Fabry-P\'{e}rot interferometer}

The UAGS spectra showed that the H$\beta$ line from the ring is
fairly well-defined and not confused with other spectral features.
It is therefore beneficial to attempt a two-dimensional kinematic
mapping with field spectroscopy.

Observations were performed during the night of 26-27 October 2008
with the scanning Fabry--P\'{e}rot Interferometer (FPI) mounted within the SCORPIO focal reducer
(Afanasiev \& Moiseev 2005) at the prime focus of the SAO RAS 6-m telescope. The desired
spectral interval in the neighbourhood of
the redshifted H$\beta$ line was selected by a narrow-band filter with full-width at half maximum (FWHM) of 21\AA\,.
The free spectral interval between the neighbouring interference orders was
18\,\AA\, ($\sim$1100 km/s). The FPI resolution,
defined as the $FWHM$ of the instrumental profile, was $1.5$\AA\,
($\sim90$ km/s) for a  0.55\AA\, sampling. The detector was a
$2048\times2048$ EEV 42-40 CCD operating in on-chip binned
$4\times4$ pixel mode to reduce readout time and match
the resultant pixel to the seeing size. The final image scale and field of
view were 0\farcs7/pixel and 6\farcm1$\times$6\farcm1, respectively.

We collected 32 successive interferograms of the object with
different spacings of the  FPI plates. The total exposure was 96~min and the seeing during
the observations varied from  1\farcs5 to 2\farcs0.  The observations were reduced  using
an IDL-based software package described by Moiseev (2002) and Moiseev  \& Egorov (2008).
Following the primary reduction, night-sky line subtraction, and wavelength calibration,
the frames were combined into a data cube where each pixel in the $512\times512$ field contains
a 32-channel spectrum centered on the H$\beta$ line sampled at 0.55\AA\ per bin. The  final
angular resolution after optimal smoothing matches the  2\farcs4 seeing.

\begin{figure*}
\centerline{
\includegraphics[width=9cm]{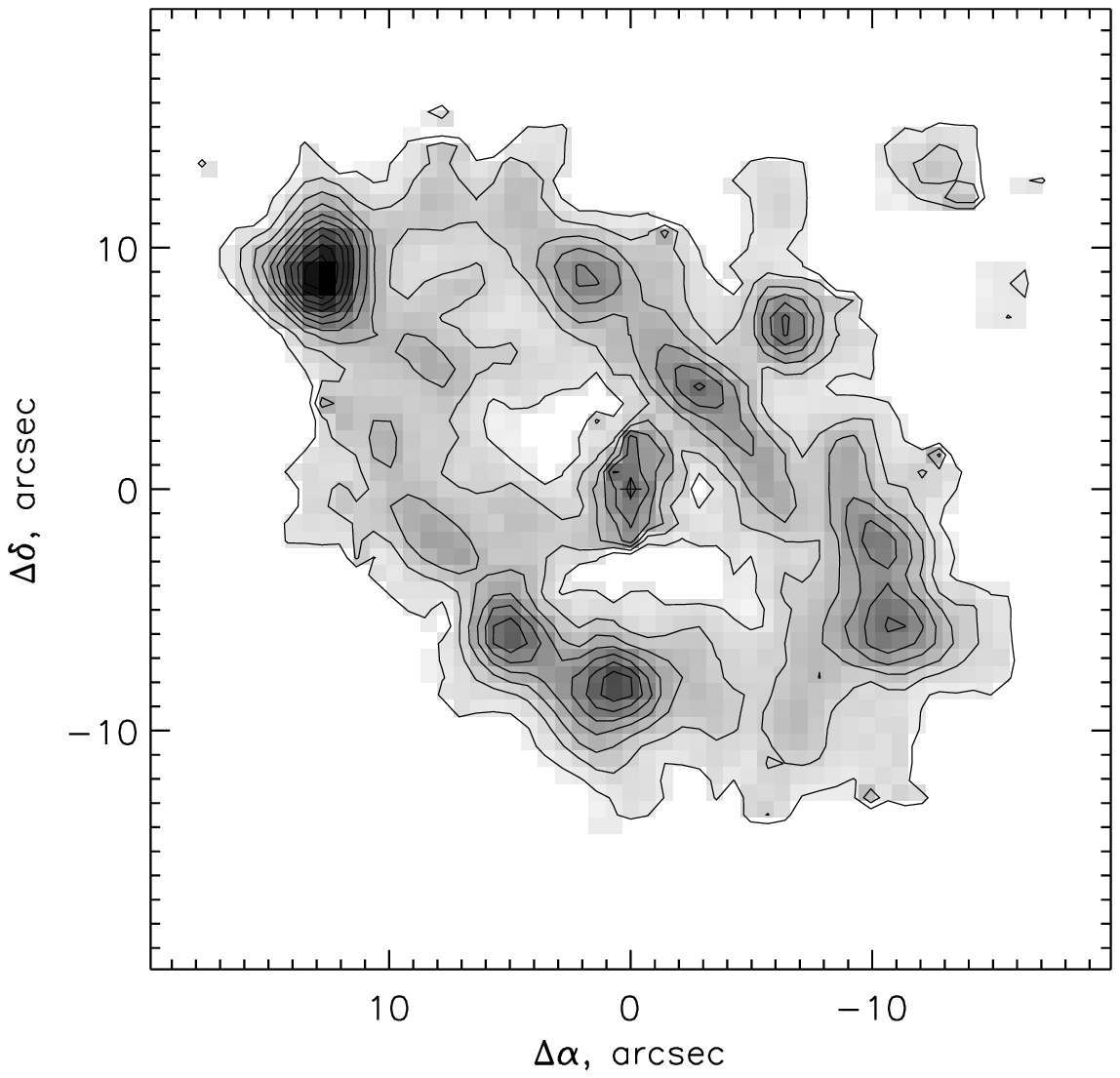}
\includegraphics[width=9cm]{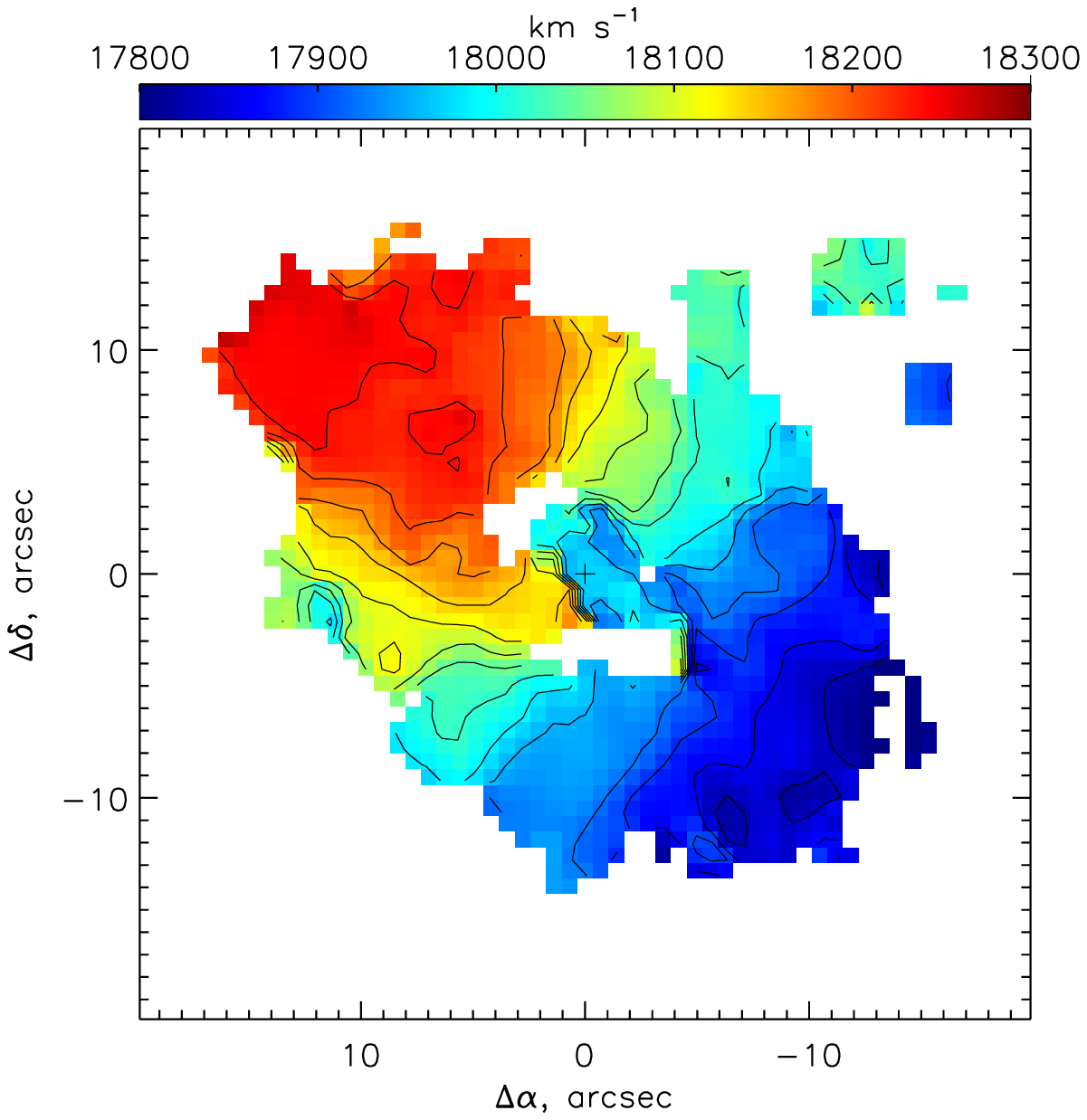}}
\caption{Results of FPI observations: intensity distribution of the H$\beta$ line emission
(left panel, with a linear intensity scale) and line-of-sight velocity field of the ionized gas (right panel).
The coordinate origin coincides with the dynamical centre of the object.}
\label{fig:FPIimage}
\end{figure*}

Figure~\ref{fig:FPIimage} shows an image of the object in the H$\beta$ line (left panel)
obtained by Gaussian fitting the line emission in the FPI data cube, and subtracting
a properly scaled combination of the line-free channels that contain the continuum contribution.
The right panel of Figure~\ref{fig:FPIimage} shows the distribution of radial velocities over
the same image. Note that there is line emission in the central object and in the bridge
connecting it to the ring. Note also the enhancements in the ring itself; these features
shall be discussed below. Figure~\ref{fig:FPIimage} shows also an emission patch $\sim$10\arcsec\
away from the ring at (--$\Delta\alpha$, $\Delta\delta$)=(13, 13); this corresponds
to a faint patch in the SDSS $gri$ deep image (see Figure~\ref{Fig:RG1_aper}).

\section{Results}
\label{txt:results}

Below we describe first the
galaxy's neighbourhood, then derive some morphological
characteristics from deep SDSS and FPI H$\beta$ imaging and surface photometry, and
finally describe the results from the spectroscopic observations.

\subsection{Neighbourhood}
\label{s:neighbourhood}

A consideration of the neighbourhood of the galaxy is relevant,
since some of the scenarios to explain the ring formation involve
the transfer of matter from a donor galaxy that survives the
transfer and could be detected, or the presence in the immediate neighbourhood of a possible
impactor galaxy.
Another consideration is the  galaxy density in the surroundings; if this is too high a polar ring
could not be expected to survive its formation for long, since it would be disrupted
 by tidal interactions from nearby objects, as already mentioned above. This would
probably be the case if a PRG would form within a dense cluster or enter one. Simulations of galaxy evolution
in clusters (e.g., Tonnesen et al. 2007) showed that complete gas  removal occurs in $\sim$1 Gyr,
while galaxies in the field, more than 2.4 Mpc from the cluster centre, often accrete cool gas.
Also, Duc \& Bournaud (2008) modelled a high-velocity encounter in a simulated Virgo cluster and
managed to reproduce the long HI tail of NGC 4254 as a result of a flyby at 1100 km s$^{-1}$
of the galaxy by a massive companion some 750 Myr ago. In the context of the galaxy neighbourhood character,
Brocca et al. (1997) did not find significant differences in the projected galaxy density between PRGs
and ``normal'' galaxies.
The characteristics of the neighbourhood of RG1  were derived from public data bases and are described below.

Given its recession velocity of 18032$\pm$47 km s$^{-1}$ and position at
07$^h$52$^m$34$^s$.4 +29$^{\circ}$20\arcmin50\arcsec, 
 RG1 is a possible member of the Abell 602 galaxy cluster at 18557
km s$^{-1}$ and is at par with the most distant PRC object. The
galaxy does not reside in the central part of the cluster, but
some 10 arcmin away from it, while the cluster diameter (adopted from NED)
is 44 arcmin. For H$_0$=72 km s$^{-1}$ Mpc$^{-1}$, its distance
is 250 Mpc and the plate scale is 1\arcsec=1.2 kpc. The projected distance from the listed cluster
centre is less than 1 Mpc. The cluster itself
was surveyed by the ROSAT all-sky survey combined with the Sloan
Digital Sky Survey (SDSS) by Popesso et al. (2004). It is a 3.4
keV cluster with an X-ray luminosity of 1.14$\times 10^{44}$ erg
s$^{-1}$ (Ebeling et al. 1998). These data suggest that RG1 may be rather close to the
cluster centre, only about half the cluster radius in projected distance, but without
any indication of the depth or of the past history of the galaxy's trajectory.

We checked the distribution of galaxies within 2 Mpc (27\farcm8 for
H$_0$= 72 km s$^{-1}$ Mpc$^{-1}$) and 500 km s$^{-1}$ of the
cluster centre using NED, to investigate the location with respect to the Abell 602 cluster.
NED lists 40 other galaxies within this volume with a median redshift of $\sim$18150 km s$^{-1}$, indicating that
this is indeed a relatively dense region of the Universe. 
RG1 has a neighbour object, SDSS J075235.90+292154.5, a compact galaxy at
18270$\pm$22 km s$^{-1}$, just $\sim$240 km s$^{-1}$ from the
redshift of RG1 and 40 arcsec North of the kink in the ring. This angular separation
translates to a projected distance of $\sim$79 kpc. The neighbour galaxy is a $g$=18.28 mag
object with H$\alpha$, [NII], [SII], [OIII], and [OII] in emission, but with the higher Balmer
lines in absorption. 

Four arcmin to the North-East, a somewhat more distant object is SDSS J075245.54+292501.9, a $g$=17.16 disky galaxy showing an absorption spectrum with a redshift of 18360$\pm$60 km s$^{-1}$. At even larger distances there are galaxies that belong to the Abell 602 cluster.


\begin{figure*}
\centerline{
\includegraphics[width=9 cm]{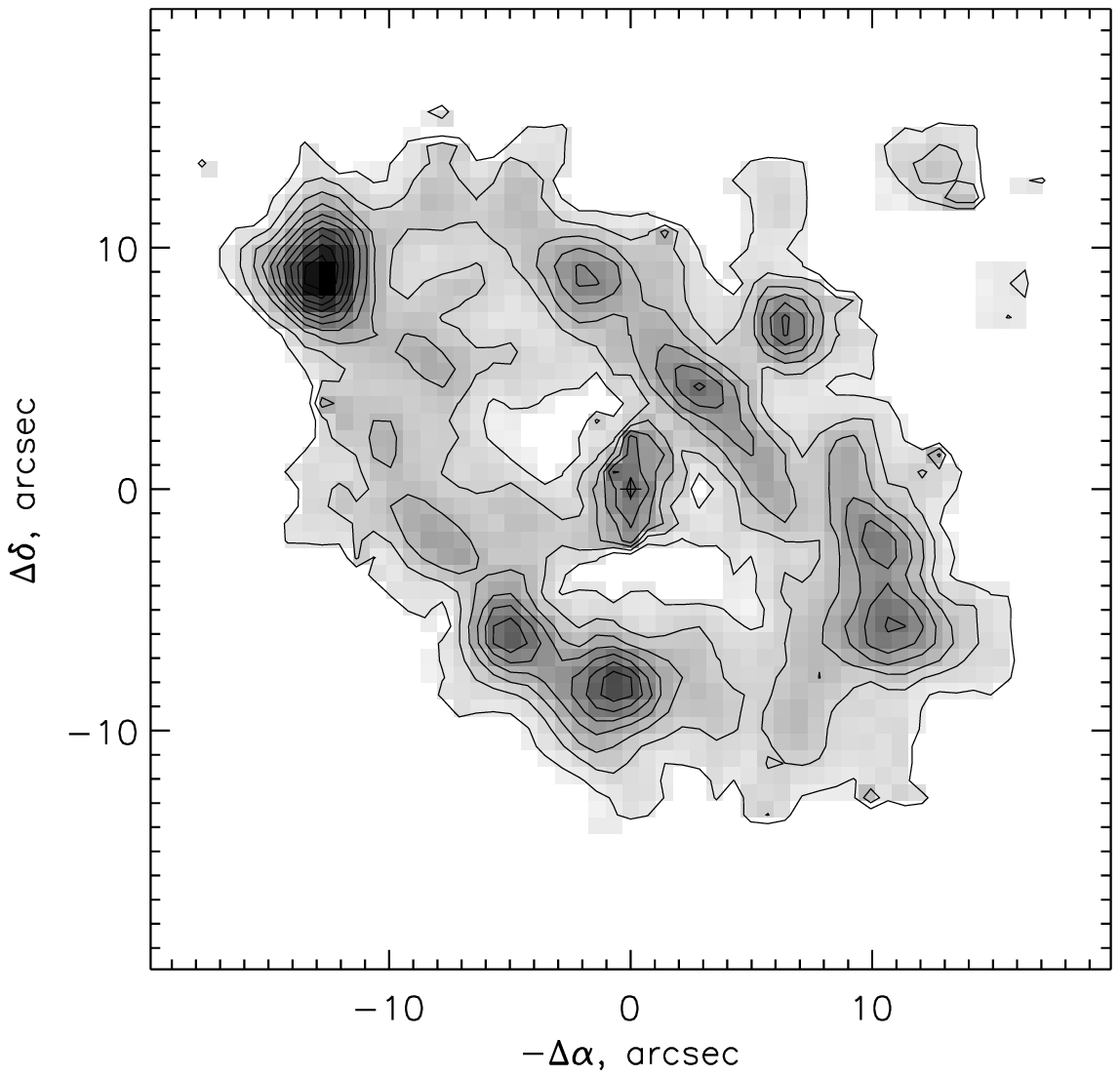}
\includegraphics[width=9 cm]{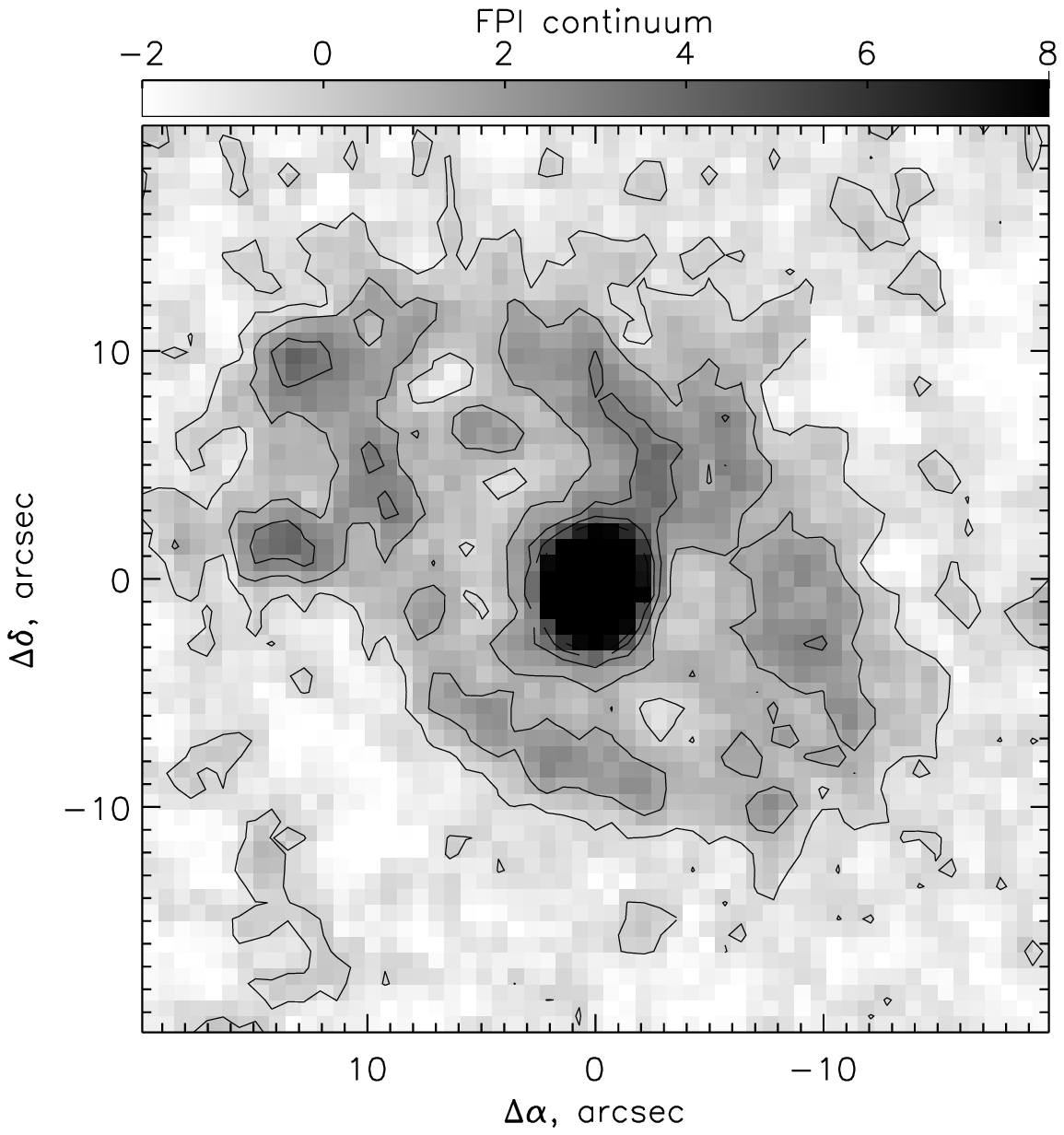}}
\caption{Results of FPI observations: intensity
distribution in the H$\beta$ emission line (left panel) and intensity of the blue continuum 
near H$\beta$ (right panel). {\bf The image in the left panel is copied from the left panel of Figure~\ref{fig:FPIimage} to allow a comparison of the line and continuum distributions.} Note the appearance of the ring in the continuum, which is very similar to that in the line emission except for the extension to the South-West.}
\label{fig:FPI_Hb_Cont_image}
\end{figure*}

\subsection{Morphological features}

The combined SDSS $gri$ image and the individual images show
that RG1 has a lenticular central
body and a disturbed ring that extends as far as
$\sim$40 arcsec$\simeq$50 kpc (major axis size) at the lowest detectable levels.
We show below that the central body fits a S\'{e}rsic brightness profile.

Specifically, the only part of the ring that appears undisturbed is its Western edge. To
the North and West the ring appears double or extended, with a fainter region between the
two branches, each of which is as wide as the single Eastern branch. The Northern region
shows a strange ``kink''. Fig.~\ref{Fig:RG1_aper} shows to the East a diffuse object
classified as ``galaxy'' in SDSS, which we call ``galaxy E''. The other diffuse object
in Fig.~\ref{Fig:RG1_aper}, which we call ``galaxy W'',
is significantly fainter than E and is located diametrically
opposite E with respect to the central body of the galaxy but is
more distant. Galaxy E is located near a cusp in the ring where the
apparent loop or kink that is rising out of the ring to the North
and East reconnects with the ring. One possibility we checked with spectroscopic observations
and report below was that this peculiar morphology could be related to the presence of galaxies E and/or W.

Figure~\ref{Fig:RG1_aper} shows a number of stellar-shaped bright
regions within the ring, mainly on its West branch. Similar bright
regions are visible in the diametrically-opposite part of the ring
near the end of the loop. We suspect that all those regions are
HII regions or young star clusters, as seen in many other ring
galaxies. 
 We confirm that this is the case based on the H$\beta$ emission line image.

We mentioned above that the central body (CB) has an elliptical or
lenticular appearance. We checked the deeper SDSS image, obtained from the $gri$ images
combination, to evaluate the nature
of the object. The CB's major axis appears connected to the
narrow part of the ring by a very faint, diffuse, linear feature. At the two
locations where this linear feature meets the ring there are two
diametrically opposite bright knots, both located on the outer
edge of the ring. These, and the linear feature, are seen also in the H$\beta$
image, standing out better in the emission line image (Figure~\ref{fig:FPI_Hb_Cont_image})
than in the continuum image. The locations
and descriptions of these features
are detailed in Table~\ref{t:RG1_details}. Columns two and three of the table give the
fractional degree coordinates of each feature.

\begin{table*}
\caption{Notable features near RG1 on the combined SDSS image and their listed SDSS photometry}
\label{t:RG1_details}
\begin{tabular}{ccccccccc} \hline
  Object  & $\alpha$ (J2000, 118$^{\circ}$) & $\delta$ (J2000, 29$^{\circ}$) & $u$  &  $g$ & $r$ & $i$ & $z$ & Description    \\ \hline
Centre & .14304 & .34717 & 19.96 & 18.14 & 17.24 & 16.80 &
16.46 & CB \\
  E  & .14752 & .34769 & 21.24 & 20.79 & 19.35 & 18.81 & 18.25 & Galaxy \\
  W  & .13634 & .34672 & 22.78 & 21.85 & 20.55 & 20.07 & 19.81 & Galaxy \\
  NW knot  & .14099 & .34915 & 20.28 &  20.19 & 20.14 & 19.95 & 19.67 & Bright region \\
  SE knot  & .14459 & .34561 & 21.96 & 22.66 & 22.93 & 22.00 & 21.78 & Bright region \\
\hline
\end{tabular}
\end{table*}
\subsection{Surface photometry using the SDSS}

\begin{figure}
{\centering
 \includegraphics[clip=,angle=-90,width=8.5cm]{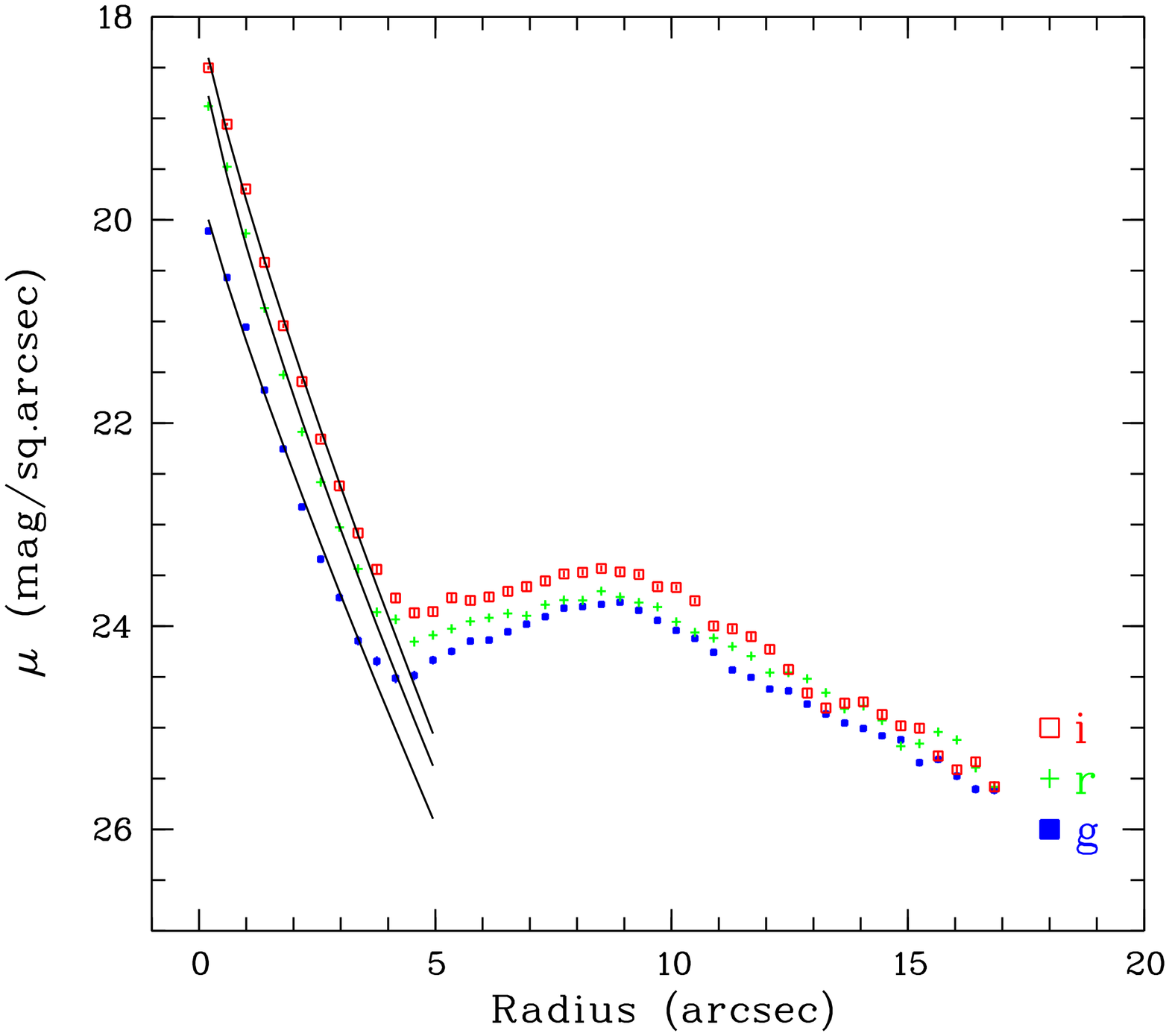}
 \includegraphics[clip=,angle=-90,width=8.5cm]{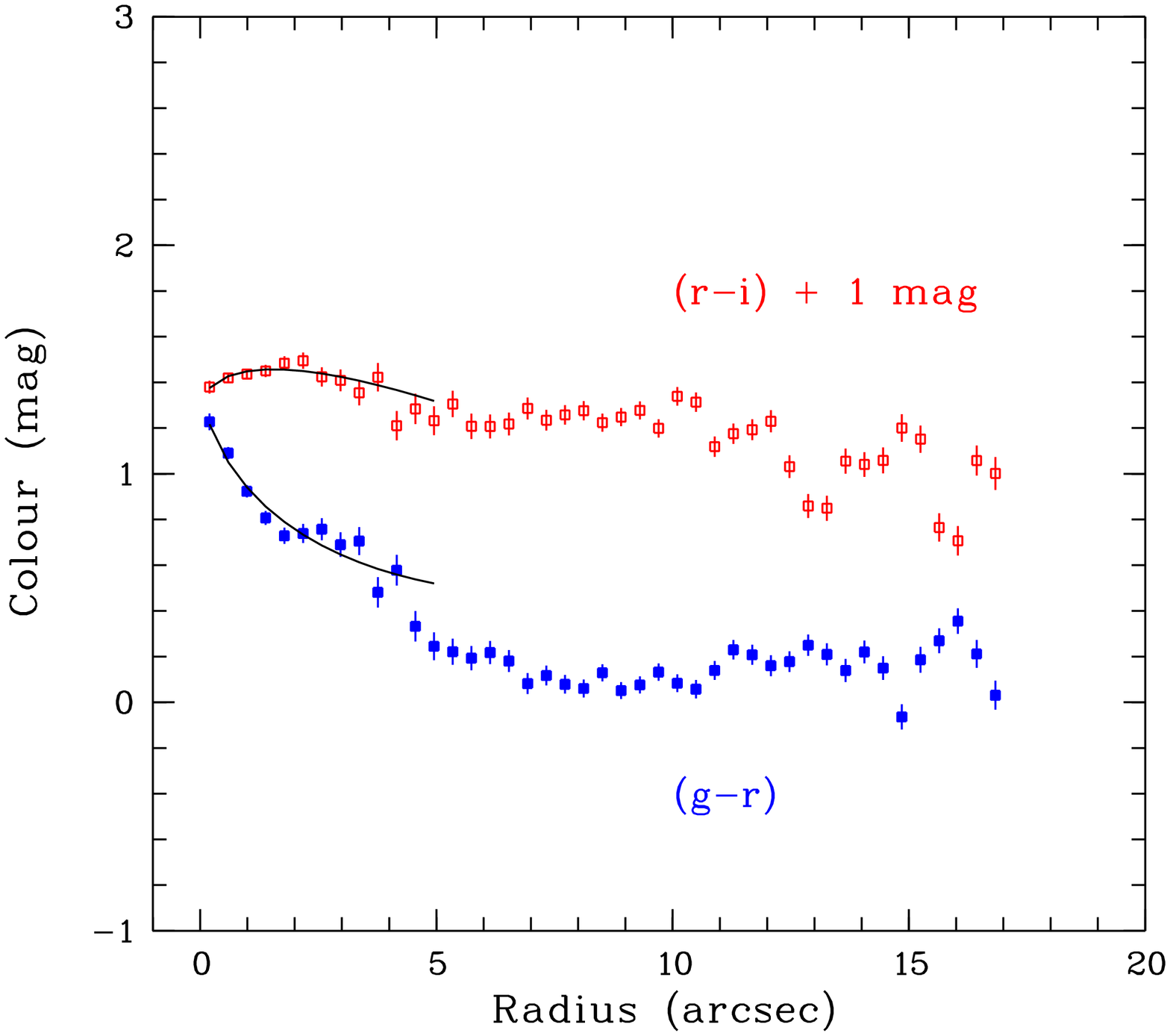}
}
 \caption{%
{\it Top panel:}
RG1 intensity profiles from the SDSS images.
S\'{e}rsic profiles have been fitted to the central component of the
galaxy up to a galacto-centric distance of four arcsec.
All the points plotted here carry  calculated
error bars derived by assuming Poisson statistics.
{\it Bottom panel:}
Colour profiles of the RG1 galaxy.
The solid line indicates the colour run for the central
object derived from the analytical S\'{e}rsic profile fits.
    \label{fig:RG1_profile_ug}}
\end{figure}

The SDSS images of RG1 in the different bands were used to produce
average surface brightness profiles that were fitted with a
S\'{e}rsic profile (Davies et al. 1988) for the central component,
after subtracting the images of the two very nearby galaxies E \& W.
The fits are shown in Fig.~\ref{fig:RG1_profile_ug} 
for the $g, r$, and $i$  SDSS bands with the fit being valid
out to four arcsec from the centre. 
 For fitting and for
cosmetic reasons we used the pixel-to-pixel information (oversampled at 0.4
arcsec pixel$^{-1}$). 
The fits show that the central component
has the profile of an elliptical galaxy, or of a galaxy bulge, following the
S\'{e}rsic surface brightness distribution.


The regions outside the central $\sim$8\arcsec$\simeq$10 kpc do not show a S\'{e}rsic
brightness distribution. The $u, g, r,$
and $i$ bands show an apparent increase in brightness, e.g., from
$\mu_g \simeq$24.5 mag/sq.arcsec at R$\simeq$4 arcsec to $\mu_g
\simeq$23.5 mag/sq.arcsec at R$\simeq$9 arcsec. This is followed by a
steady decrease in surface brightness. The $z$-band image shows an approximately constant
surface brightness from R$\simeq$4 arcsec to $\sim$12 arcsec, followed by a decrease in $\mu_z$.

Using multi-aperture photometry, we produced colour profiles from the SDSS images.
Some are shown in the lower panels of Figure~\ref{fig:RG1_profile_ug}.
The figure also shows the run of the colour with
the S\'{e}rsic profiles fitted for the intensity images in all filters.
The solid lines  representing the
analytical fit match reasonably well points closer that $\sim$5
arcsec from the centre, but not those further out.

We used the irregular mask derived from the deep combined $gri$ image
shown in Figure~\ref{Fig:RG1_aper} to perform surface photometry on the
 SDSS $ugriz$ images. The total counts were added within this irregular aperture and the
sky contribution from an equivalent area of the sky was
subtracted.
The total photometry results through this irregular
aperture that includes all the parts of the ring were used to calculate the
light contribution from the ring itself, by subtracting the light
from the central object as given in SDSS from the measurement for
the entire object. The results are given in Table~\ref{t:photom}
and show that the ring is blue while the central body is reddish. The colours calculated
here match well those derived from the profiles of the galaxy,
shown in the lower panel of Figure~\ref{fig:RG1_profile_ug}.
The ring produces about the same
amount of light as the central object in the $i$ band, but is significantly
brighter at bluer bands. The ring is also significantly bluer than the central object,
with e.g., ($g-r$)=0.14 mag, vs. 0.87 mag for the central object.

\begin{table*}
\caption{Photometry of the main RG1 components} \label{t:photom}
\begin{tabular}{cccccc} \hline
  Object/Band         & $u$  & $g$ & $r$   &  $i$ & $z$    \\ \hline
Total                 & 17.96$\pm$0.06 & 16.80$\pm$0.01 & 16.41$\pm$0.01 & 16.12$\pm$0.01 & 16.00$\pm$0.04 \\
CB        & 20.15$\pm$0.03 & 18.22$\pm$0.01 & 17.35$\pm$0.01 & 16.92$\pm$0.01 & 16.59$\pm$0.01 \\
Ring (difference)     & 18.12$\pm$0.07 & 17.14$\pm$0.01 & 17.00$\pm$0.02 & 17.01$\pm$0.01 & 16.94$\pm$0.04 \\
           \hline
\end{tabular}

\end{table*}

The enhanced brightness at R$\simeq$9 arcsec, and the steady
decrease in intensity further out, are even more accentuated in
the $u$ band and are almost invisible in $z$. The longest wavelength
band shows the ring at a steady surface brightness of $\mu_z
\simeq$23.5 mag/sq.arcsec. However, this enhanced surface brightness is not
reflected in the colour plots, which stay approximately constant
from 5 arcsec to 14 arcsec. We note that in this procedure we did not try to
eliminate the light contributed by the bright knots in the ring; these might possibly
be star clusters and/or HII regions, and might have different colours than the general population of the ring.
Given that the distance modulus to the object of 37.15 mag, the central object has
M$_r$=--20.25 and the ring has M$_r$=--19.85.



\subsection{Long-slit Spectroscopy}
\label{s:long_slit}


The 2D spectrum taken with 6m BTA telescope in 2001 (shown in
Figure~\ref{Fig:2D-spectrum}) covers a wider wavelength range than the two SCORPIO spectra
and can provide a perspective view of the object. For this purpose, the spectrum is discussed below in qualitative terms.
The spectrum shows the signature of an evolved
stellar population for the central object with Ca II and G-band
absorption, though the [OII] emission lines are visible as a
blended feature not only at the very centre (and seen both on the
SDSS spectrum and on that extracted for the central region from
our data, as explained below) but extending some two arcsec on
either side of the centre. Therefore,
the [OII] emission is probably not restricted to the very centre of the
galaxy but extends slightly beyond it. This extended line emission from the central object
is supported by the detection of H$\beta$ emission from the same location (see Figure~\ref{fig:FPIimage}).

The strong absorption lines of Ca II and the CH G-band do not seem to be significantly
inclined and do not indicate a fast rotation of the stellar
component. The spectral resolution is not sufficient to derive
line widths in order to determine the run of the velocity
dispersion with galacto-centric distance in the central body.

The spectrum shows a number of interesting features. The central
object yields an extracted spectrum that is virtually identical
with that of SDSS, which was obtained with a three-arcsec wide fiber
placed on the galaxy's brightest spot. The long slit allows the
observation of the spatially-resolved ring spectrum. This shows
emission lines from [SII], [NII], H$\alpha$, possibly [OI],
[OIII]$\lambda\lambda$5007, 4959\AA\,, H$\beta$, H$\gamma$, and
[OII]$\lambda$3727\AA\, doublet blended. The underlying continuum of the
ring is blueward-sloping and the SW part of the spectrum is
fainter than the NE part.

We estimate that the NE part of the slit shows stronger line
emission than the SW part. The emission between the central object
and the ring increases steadily from $\sim$3 arcsec to $\sim$12
arcsec, then drops to a level where it cannot be measured anymore
18 arcsec away from the centre. The width of the
H$\alpha$-emitting region in the NE part of the ring is $\sim$8
arcsec$\simeq$10 kpc.

The opposite side of the ring, to the SW, shows a much fainter H$\alpha$ line that extends
from $\sim$6 arcsec to $\sim$14 arcsec from the centre with a constant intensity, covering also $\sim$10 kpc. The kinematic behaviour of this side is, as already
mentioned, very different from that of the other side; the gas shows essentially the same
recession velocity irrespective of distance.



\begin{figure}
{\centering
 \includegraphics[clip=,angle=0,width=8.0cm]{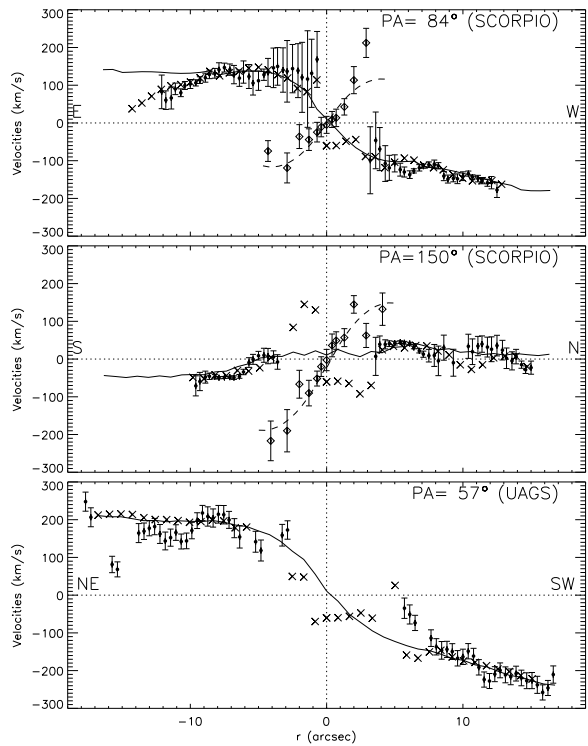}
}
 \caption{%
Rotation curves of RG1 for the ionized gas and for the stars. The specific slit
position angle and the approximate directions are indicated, as is the
instrument that acquired the spectrum. The scale is 1\arcsec=1.2 kpc.
The recession velocity of the H$\alpha$ emission line
(filled  circles) and the Na I absorption line (empty rhombi) are plotted, along with
the corresponding cross-sections through the FPI H$\beta$ velocity
field (cross symbols) discussed below.
The solid lines mark the
cross-section of the model velocity field derived from FPI data. The
thick dashed line is a projection of the mean rotation curve for the
stellar component.
    \label{fig:Ring_RG1_rot}}
\end{figure}

The spectra obtained in 2008 with SCOPIO are superior in quality to the UAGS spectrum and allow a
better derivation of the kinematic properties of the stars and of the gas in the galaxy. 
We derived  velocity-distance plots from the 2001 and 2008 BTA spectra for the H$\alpha$
and the NaI lines, which are shown in Figure~\ref{fig:Ring_RG1_rot}.

\section{Interpretation}
\label{txt:interp}

This paper started by describing an object that, at first look, appears to be a ring galaxy.
We presented above observational evidence originating from SDSS imaging, and from long-slit
and Fabry-P\'{e}rot spectroscopy at the 6-m BTA telescope, in an attempt to understand the true nature of the object.

At this point, the questions we attempt to answer are:
\begin{enumerate}

\item What is the nature of the central object in the galaxy? Could this be the ``projectile''
that produced a collisional ring galaxy?

\item Are the two nearby galaxies in the RG1 system, objects E and W, involved in
the formation and evolution of the ring galaxy?

\item What is the nature of the ring ? Could it be the outcome of a collision? Are there any
signs of accretion that could have produced the ring by another mechanism?

\item Is there a physical connection between the central object and the ring around it?

\item What is the nature of the ionized gas seen in the object? Is this a gas disk ring, or
is it a disk, albeit with a warp?


\end{enumerate}

The answer to the second question can already be given, using the UAGS and SCORPIO
long-slit spectroscopy.
The spectra of galaxies E and W 
indicate that these are not related to RG1 but are seen in projection.
On the other hand, the RG1 spectra yield a wealth of information about the internal kinematics of the object.
In the following sections we describe sequentially the conclusions derived from the long-slit spectroscopy,
from FPI observations, and 
from the surface photometry.

\begin{figure}
\includegraphics[width=8.0cm]{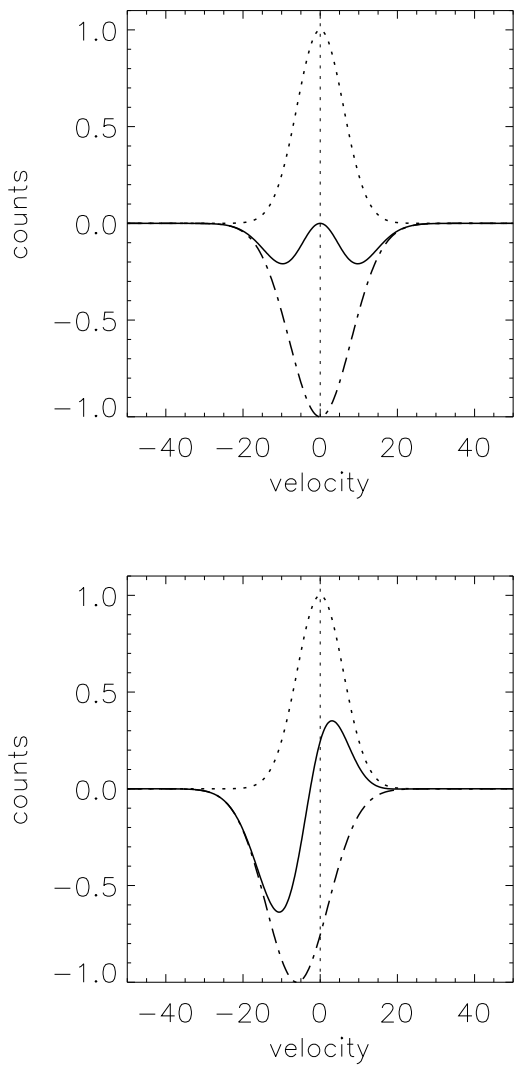}
\caption{Combining absorption and emission components to produce the H$\beta$ profile. The
dotted and dash-dotted Gaussians correspond to emission and absorbtion components and the
solid line is their sum, thus the total (observed) profile.  The   velocity of the emission
component (``gas'') is zero, and the velocity of the absorption component (``stars'') is either zero
(top panel) or -7 km s$^{-1}$ (bottom panel).}
\label{fig3}
\end{figure}

\subsection{Slit spectra and FPI data}


Our long-slit UAGS and SCORPIO
spectroscopy 
shows that the two diametrically-opposite parts of the ring show the most intense and high-excitation emission. At these ends of the ring we see intense [OIII] emission; this might be present elsewhere in the object but it must be of much lower intensity and equivalent width. 
The velocity-distance plots are presented in Figure~\ref{fig:Ring_RG1_rot}. We estimated qualitatively the orientation parameters of the stellar and gaseous components of RG1 by deprojecting the kinematic profiles assuming a pure disk rotation model, where the rotation curve is described by a second-order polynomial.






The SCORPIO and UAGS observations match each other very well. We modeled the stellar rotation curve with a second degree polynomial and obtained a PA=--53$^{\circ}\pm$4$^{\circ}$ and an inclination of i=50$^{\circ}\pm$15$^{\circ}$, while for the emission lines we measured a PA=59$^{\circ}\pm$6$^{\circ}$ and an inclination of 61$^{\circ}\pm$10$^{\circ}$. For the gaseous ring we 
must adopt different shapes for the rotation curves on the NE and SW sides of RG1 since, as Figure~\ref{fig:Ring_RG1_rot} shows, the rotation curve seems to be asymmetric. To approximate it, we adopt the mean of the second degree polynomial fit to the SW side and that fitted to the NE side.

The large angular coverage of the FPI kinematic data argues that this should be the preferred source
from which to derive the kinematics of the ionized gas.
The line-of-sight velocity field of ionized gas and the H$\beta$ monochromatic map
shown in Figure~\ref{fig:FPIimage} were constructed by fitting Gaussians to the emission line
spectra in the data cube. 
 The velocity field shows a regular pattern of rotation at distances $r\geq$3-4\arcsec\
from the nucleus. The velocities measured for H$\beta$ with the FPI match very well
the previous long-slit H$\alpha$ data (see crosses plotted in the three panels of
Figure~\ref{fig:Ring_RG1_rot}). The exceptions are two
points at $r\approx-15''$ for $PA=58^\circ$, 
probably caused by an error in UAGS measurements. The H$\beta$ velocity behaviour
changes significantly in the   inner ($r<4''$) region. A strong velocity gradient in
the stellar absorption components
along the minor axis is present there (see Figure~\ref{fig:Ring_RG1_rot} for $PA=150^{\circ}$).

\begin{figure*}
\centerline{
\includegraphics[width=9 cm]{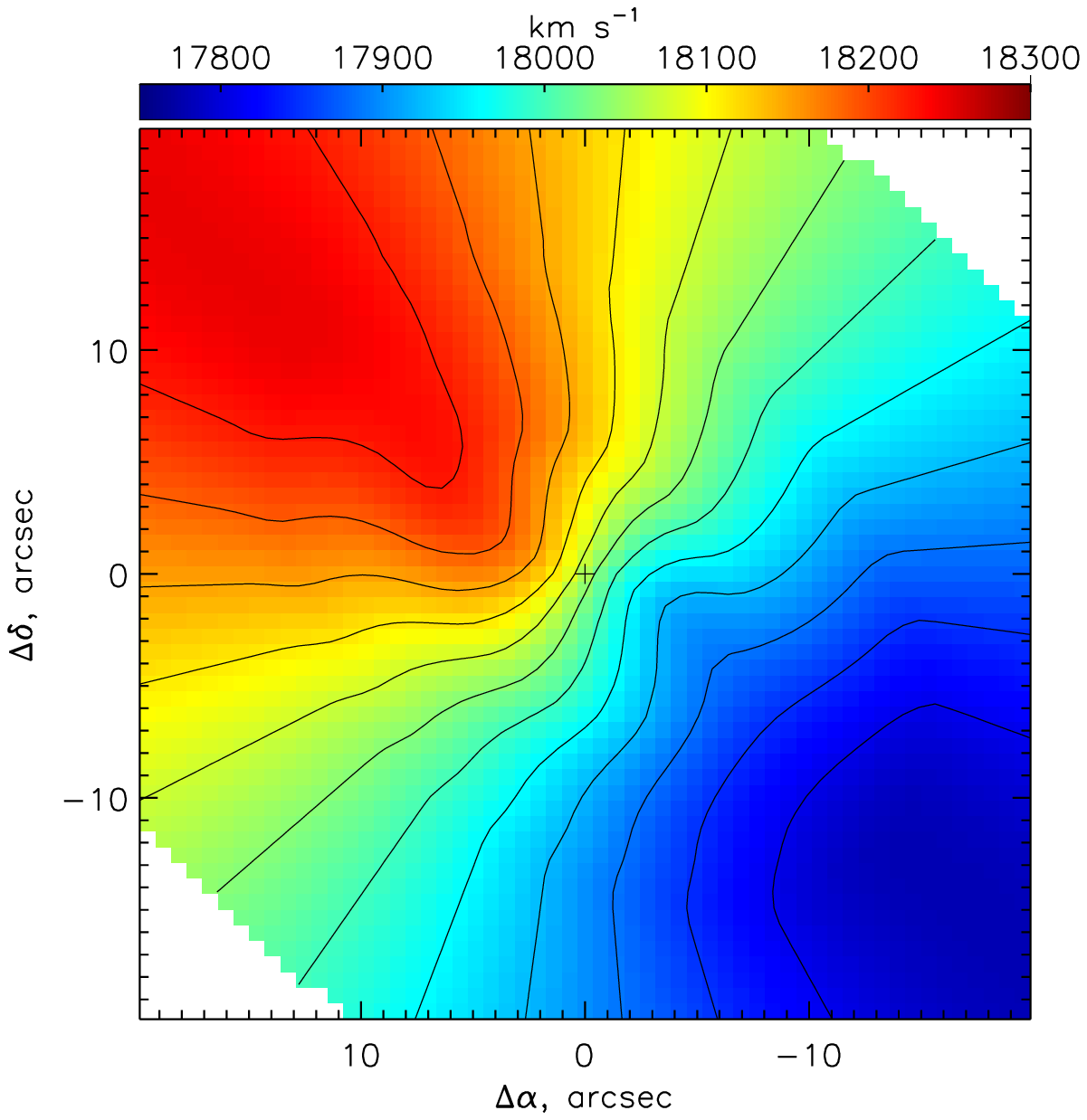}
\includegraphics[width=9 cm]{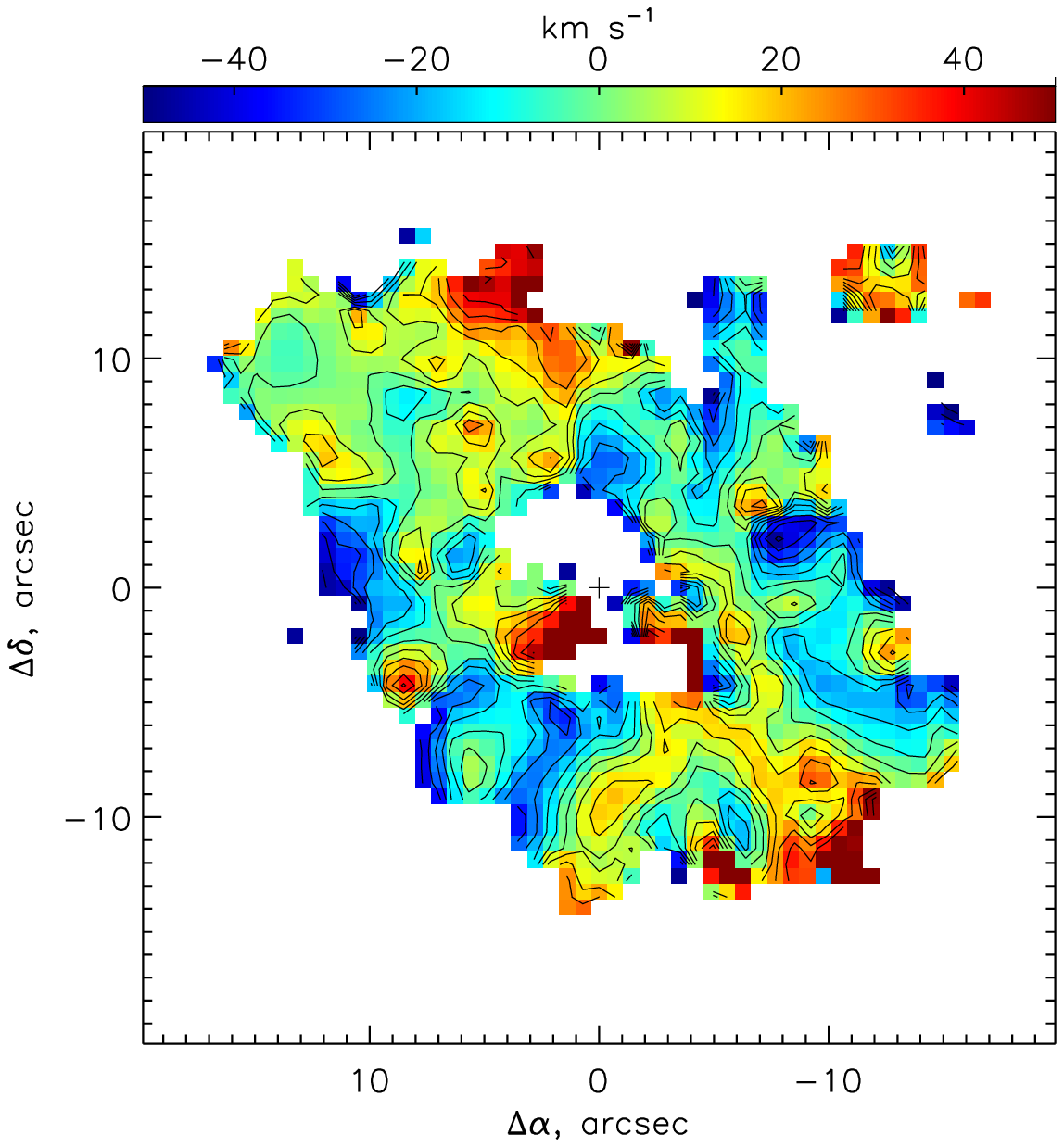}}
\caption{Kinematic information derived from the FPI H$\beta$ observations. Velocity
field for the quasi-circular rotation model (left panel)
and map of residual velocities (right panel). }
\label{fig5}
\end{figure*}

In the long-slit data we cannot be sure of the detection of the H$\alpha$ emission because of 
possible contamination by underlying stellar absorption lines. In addition, this region is possibly
affected  by the residual fringe pattern of the CCD chip. In contrast with the H$\alpha$ data,
the FPI allows a confident detection of the broad $\sim$200 km s$^{-1}$ $H\beta$ emission
line at the center of the galaxy.
Surprising, the $H\beta$ kinematics seem to indicate counter-rotation relative to the stellar
velocity gradient. 
We suggest that this is not the real situation
and propose an alternative solution.

For demonstration purposes, we assume that the $H\beta$ emission
component  detected in FPI data originates from the ionized gas and the absorption component from stellar photospheres produce the same H$\beta$ equivalent widths and have similar velocity dispersions.
We stress that the discussion to the end of this paragraph relates only to the FPI H$\beta$ data.
In the nucleus we observe a combination of an emission line and a broad H$\beta$ stellar absorption
(Figure~\ref{fig3}). If the stars and ionized gas would have the same velocities ($v=0$, plotted
in the top panel of Figure~\ref{fig3}), then the peak of their summed profile would also have the
same  velocity. However, if the absorption component would be blueshifted while the emission would
remain at zero velocity, then the summed components would show a redshifted peak velocity, because
the negative contribution of the absorption line wing will be smaller under the red part of the
emission line   (bottom panel of Figure~\ref{fig3}). 

The relatively small free spectral range of the FPI, the interfringe spacing, makes it impossible to correctly
remove the continuum contribution. Moreover, overlapping of neighbouring interference orders amplifies the
effect of the ``slanting continuum''. It is possible that the ionized gas rotation in the inner region
follows the same behaviour as at the largest radii: a zero gradient along the minor axis (PA=140$^{\circ}$-150$^{\circ}$),
redshifted to the NE and blueshifted  to the SW-side of the gaseous ring.

\subsection{A quasi-circular rotation model}

Below we present a simple model that explains the main kinematic properties of the object.
The location of the centre of rotation was calculated under the assumption
that  the velocity distribution for distances $r$=6--20\arcsec is symmetric. The position
of the kinematic centre agrees within $\pm$1.5\arcsec\ with the centre of the continuum and
H$\beta$ images. We fixed the rotation centre at the optical nucleus for the  analysis described below using a symmetry assumption. The kinematic behaviour was analyzed
using the ``tilted rings'' method  (Begeman 1989). The  velocity field was divided
into one-arcsec-wide elliptical rings oriented according to the adopted position angle of
the major axis (PA$_0$) and inclination of the disk ($i_0$).

\begin{figure}
\centerline{
\includegraphics[width=8 cm]{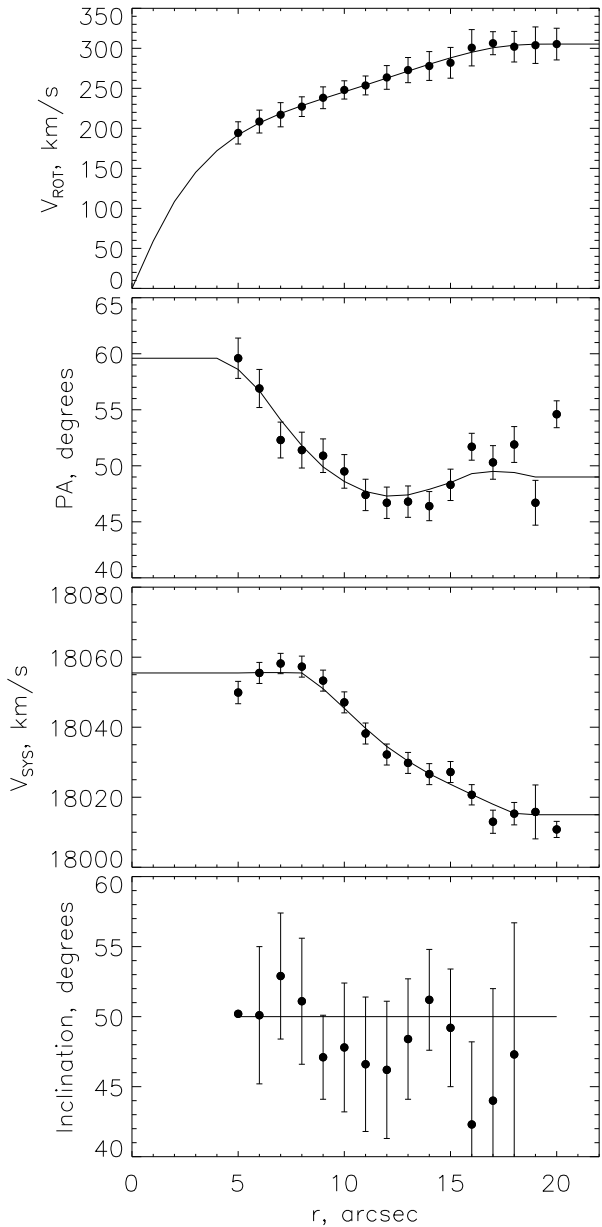}}
\caption{Radial variation  of kinematic parameters characterizing the velocity field of
ionized gas (from top to bottom): circular rotation velocity, position angle, systemic velocity,
and inclination. The filled circles  correspond
to the tilted-rings measurement. The solid line corresponds to the parameters of the two-dimensional
model shown in the Figure~\ref{fig5}. }
\label{fig4}
\end{figure}

\begin{figure}
\includegraphics[width=8.5cm]{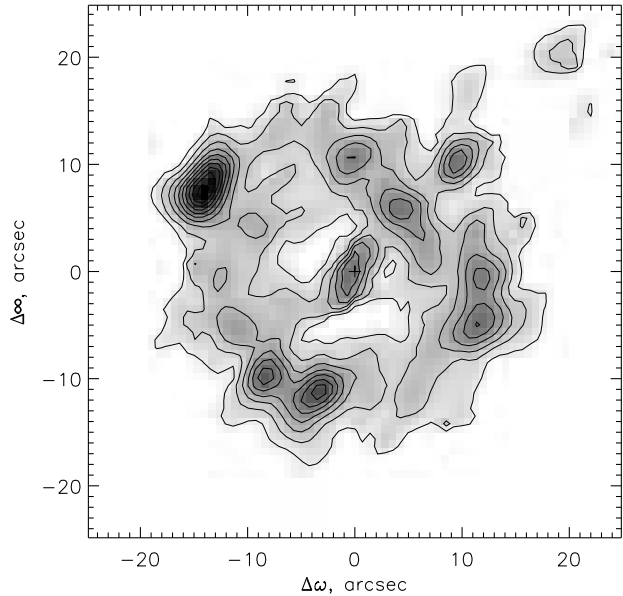}
\caption{Distribution of the H$\beta$ emission after deprojection of the polar disk plane using the FPI total line emission image.}
\label{fig6}
\end{figure}

Using the least-squares method, we best-fitted the position angle of the kinematical axis PA$_{kin}$,
the mean velocity of rotation $V_{rot}$, systemic velocity $V_{sys}$ and inclination $i$
for each ring. The method is justified in Moiseev, Vald{\' e}s \& Chavushyan  (2004) where
references to the original papers are given. The kinematic model is shown in the left panel of
Figure~\ref{fig5}; the right panel of this figure shows the residuals of this model when compared
to the real data. These residuals are mostly smaller than 20 km s$^{-1}$ and no specific
trend is visible, implying that the model fits rather well the data.

Figure~\ref{fig4} shows the radial variations
of the model parameters. The rotation curve is rising up to $r$=18\arcsec\,, then remains approximately
flat at $V_{rot}=305$ km s$^{-1}$. The  PA variations are not large, only $\sim15^\circ$ in amplitude,
and have a systematic nature that may be connected with a warp of the disk  or possibly with
the presence of density waves in the disk.
The relative errors of the individual  measurements of the inclination  make it impossible
to draw any conclusions about variations of $i$ with radius. Within the approximation of a flat
circular rotating thin disk (without a warp), we found two formal solutions for the orientation:
PA$_0=49^\circ\pm3^\circ$  with inclination $i=50^\circ\pm10^\circ$, and relative inclination $\Delta i=58^\circ\pm10^\circ$ {\bf or $\Delta i=73^\circ\pm11^\circ$}. These can be used as a mean disk orientation, and were
adopted to produce the rotation curve shown in Figure~\ref{fig4}. Note also that here, as in
other similar galaxies (e.g., Arp 212: Moiseev 2008) there are two possible solutions
for the relative inclination angle. One was listed above; the other is $\Delta i'=73^\circ\pm11^\circ$.

The third panel down of Figure~\ref{fig4} shows that there is a systematic variation of
the centroid velocity (systemic velocity) for each fitted ring. Specifically, from the inner
edge of the ring to its outer edge, the systemic velocity changes smoothly by about $50$ km s$^{-1}$; this is surprising and unusual. Such a velocity gradient in the radial direction may indicate
bulk motion of (at least) the gaseous material in the ring. Specifically, the ring material
might not only be orbiting the central object on circular trajectories, but might also be
infalling or expanding. The maximal velocity would be at the outer radius of the ring.
We observe that the radial gradient decreases the closer the material is to the central
object.

If the radial motion would be infall, then this could be a result of some braking
force in the inner parts of the ring (possibly gas pressure). If the radial motion would be outwards, the material could be accelerated away by radiation pressure (rather unlikely, given the apparent nature of the central object). A similar kinematic behaviour was observed by Kenney et al. (2004) for NGC 4522; there it was explained as the signature of stripping the outer gas layers of the galaxy by the intergalactic medium of the Virgo Cluster. However, there the outer gas layers were blueshifted to a velocity closer to that of the mean cluster redshift, whereas here the velocity difference with respect to the median redshift of the cluster objects described in \S~\ref{s:neighbourhood} increases with galacto-centric distance. This could mean that the galaxy is interacting with the intra-cluster medium by infalling while the ring material is ``peeling off'' due to an interaction with the intra-cluster material. 

Figure~\ref{fig6} shows a view of the H$\beta$ line image after deprojection for inclination.
This shows again that the ring seems to be complete, though some regions show enhanced brightness.
The central object, which shows a S\'{e}rsic light distribution profile in the SDSS bands, exhibits isophote twisting. 


\subsection{The stars}

The images of the object in the deep $gri$ combined image (Figure~\ref{Fig:RG1_aper}),
in the blue continuum, and in
the H$\beta$ line  (Fig.~\ref{fig:FPI_Hb_Cont_image}), indicate that the
central part of RG1 seems to be connected by a faint bridge to
the ring. This bridge is visible in the  ring minor axis direction but is not visible in the
perpendicular direction. This could indicate that RG1 may not be a
true ring galaxy, but rather a barred spiral galaxy with two
semi-circular arms, a weak bar and a strong flattened spheroid. 

The central body of RG1 has the appearance of an elliptical or lenticular galaxy, and
could be considered the collider galaxy in a scenario where RG1 would be a collisional
ring galaxy. Our spectroscopy shows that the central velocity dispersion for the stars is
commensurate with the maximal rotational velocity; 
 this is not typical for a disk. This interpretation, that the
CB is not a disk, is also supported by the integrated colours, derived either via the total photometry or as average ring colours from the set of circular aperture
measurements. However, the angular resolution is not sufficient to
draw any strong conclusion about the morphology, even based on a more
sophisticated variable elliptical contour fitting. The light distribution follows a
S\'{e}rsic profile with a power index of 1.1 to 1.3, depending on colour;
this might be understood as the signature of an exponential disk tilted
with respect to the line of sight, as calculated above, or as that of a rotationally-supported
oblate spheroid, or an Sa/Sb bulge. Such an interpretation is supported by the location of the object
in an ellipticity vs. v$_{\rm max}$/$\sigma_{\rm eff}$ diagram, as shown in Figure 3
of Kormendy \& Illingworth (1982). There is a hint of a slight blueing-out
in the ($g-r$) colour profile (see Fig.~\ref{fig:RG1_profile_ug}), but this effect
is less clear in the other colours.

Since the ring colours appear constant throughout the ring's
extent, this seems to indicate a uniform stellar
population type in the ring that does not change with galacto-centric
radius, with only the amount of stars per unit area changing. The light from the ring,
thus the stellar surface density, peaks at R$\simeq$9''$\approx$11 kpc from the centre.

The ring is much bluer than the central object, as our surface and integrated photometry show
(see Figure~\ref{fig:RG1_profile_ug}).
Lisker et al. (2006) used the ($g-r$) colour as an age indicator in
composite populations of dE galaxies and their method can be applied here. Their models show that for colours as blue as the RG1 ring, ($g-r$)=0.01, the young stellar
population must be some 1\% of the total stellar population and
must be less than 10$^7$ years old (top panel of their Figure 10).
Note that we did not correct the SDSS photometry results for Galactic extinction; the foreground extinction is expected to produce E(B--V)=0.049 (Schlegel et al. 1998) but this prediction could be influenced by the presence of galaxies even in low-extinction regions (Yasuda et al. 2007) implying that this E(B--V) estimate could even be smaller. We also did not apply a k-correction; this and the foreground extinction are not expected to influence significantly our conclusions given the possible influence of the internal extinction and of the emission line contamination of the SDSS colours.

The presence of
a substantial population of young stars is also supported by the detection of strong
emission lines from the ring.
Shimasaku et al. (2001) investigated the photometric colours of
456 galaxies using SDSS photometry. They give average colours and
dispersions in colours for different morphological types as
characterized by the T types in their Fig. 5 and Table 1. A
comparison of the results obtained here and the data presented by
Shimasaku et al. shows that the colours of the RG1 ring correspond
to Magellanic Irregular galaxies (type Im with T$\geq$5.5), which fits the character
of the spectrum analyzed here. The colours of the central object
of RG1 correspond to those of an elliptical galaxy although, as we showed above, emission
lines are present in the central object as well.


\begin{table*}
\centering{
\caption{Line intensities of studied areas}
\label{t:Intens}
\begin{tabular}{lcccccc} \hline
\rule{0pt}{10pt}
& \MC{2}{c}{Area 1} & \MC{2}{c}{Area 2}  & \MC{2}{c}{Area 3}   \\ \hline
\rule{0pt}{10pt}
$\lambda_{0}$(\AA) Ion                  & F($\lambda$)/F(H$\beta$)&I($\lambda$)/I(H$\beta$)
& F($\lambda$)/F(H$\beta$)&I($\lambda$)/I(H$\beta$)
\\ \hline
3727\ [O\ {\sc ii}]\                    & 7.58$\pm$2.57 &15.92$\pm$5.11  & 4.88$\pm$0.60 & 8.24$\pm$1.08 & 3.96$\pm$1.32 & 5.07$\pm$2.43 \\
4861\ H$\beta$\                         & 1.00$\pm$0.56 & 1.00$\pm$0.56  & 1.00$\pm$0.13 & 1.00$\pm$0.17 & 1.00$\pm$0.34 & 1.00$\pm$0.52 \\
4959\ [O\ {\sc iii}]\                   & 1.30$\pm$0.53 & 1.00$\pm$0.50  & 1.11$\pm$0.15 & 1.07$\pm$0.14 & 1.65$\pm$0.47 & 1.16$\pm$0.45 \\
5007\ [O\ {\sc iii}]\                   & 3.30$\pm$0.55 & 2.45$\pm$0.43  & 3.22$\pm$0.33 & 3.04$\pm$0.32 & 3.34$\pm$0.91 & 2.31$\pm$0.87 \\
6548\ [N\ {\sc ii}]\                    & ---           & ---            & 0.14$\pm$0.10 & 0.08$\pm$0.06 & ---           & ---           \\
6563\ H$\alpha$\                        & 9.34$\pm$2.38 & 2.85$\pm$0.81  & 5.08$\pm$0.51 & 2.88$\pm$0.32 & 7.05$\pm$1.74 & 2.92$\pm$1.08 \\
6584\ [N\ {\sc ii}]\                    & ---           & ---            & 0.48$\pm$0.13 & 0.27$\pm$0.08 & ---           & ---           \\
6717\ [S\ {\sc ii}]\                    & ---           & ---            & 0.98$\pm$0.14 & 0.53$\pm$0.08 & ---           & ---           \\
6731\ [S\ {\sc ii}]\                    & ---           & ---            & 0.37$\pm$0.09 & 0.20$\pm$0.05 & ---           & ---           \\
  & & \\
C(H$\beta$)\ dex                     & \MC {2}{c}{1.3$\pm$0.6}    & \MC {2}{c}{0.7$\pm$0.1}    & \MC {2}{c}{0.8$\pm$0.3}  \\
F(H$\beta$)$^a$\                     & \MC {2}{c}{1.2$\pm$0.6}    & \MC {2}{c}{4.0$\pm$0.4}    & \MC {2}{c}{2.0$\pm$0.5}    \\
EW(H$\beta$)\ \AA\                   & \MC {2}{c}{  16$\pm$ 7}    & \MC {2}{c}{  26$\pm$ 2}    & \MC {2}{c}{  10$\pm$ 2}    \\
  & & \\
$T_{\rm e}$(OIII)(K)\                & \MC {2}{c}{12,600$\pm$3850~~}   & \MC {2}{c}{10,900$\pm$1250~~}  &  \MC {2}{c}{9,450$\pm$2800~~}       \\
$T_{\rm e}$(OII)(K)\                 & \MC {2}{c}{14,100$\pm$3750~~}   & \MC {2}{c}{11,000$\pm$1350~~}  &  \MC {2}{c}{9,600$\pm$2500~~}       \\
O$^{+}$/H$^{+}$($\times$10$^5$)\     & \MC {2}{c}{17.460$\pm$18.590~~} & \MC {2}{c}{22.220$\pm$3.536~~} & \MC {2}{c}{24.480$\pm$19.020~~}     \\
O$^{++}$/H$^{+}$($\times$10$^5$)\    & \MC {2}{c}{4.530$\pm$4.473~~}   & \MC {2}{c}{8.370$\pm$3.154~~}  & \MC {2}{c}{11.550$\pm$12.760~~}     \\
12+log(O/H)\                         & \MC {2}{c}{~8.34$\pm$0.38~~}    & \MC {2}{c}{~8.49$\pm$0.07~~}   & \MC {2}{c}{~8.56$\pm$0.28~~}        \\
\hline
\MC{3}{l}{~~} \\
\MC{3}{l}{$^a$ in units of 10$^{-16}$ ergs\ s$^{-1}$cm$^{-2}$.}\\
\end{tabular}
 }
\end{table*}


\subsection{Determination of the metal abundance}
\label{txt:abund}

After comparison of the spectrum of central part with the SDSS data, and the small
additional corrections described in Section~\ref{txt:long_slit_obs},
we can use the UAGS spectrum with its large wavelength coverage to obtain
a metal abundance estimation. The two SCORPIO spectra are used for the evaluation
of the [\ion{N}{ii}]/H$\alpha$ ratio at two other position angles.
The [\ion{S}{ii}] $\lambda$6716,6731 lines are badly affected by fringing;
while these lines can be used for kinematic studies, they are useless for abundance studies.

\begin{figure}
\includegraphics[width=8.0cm,clip=]{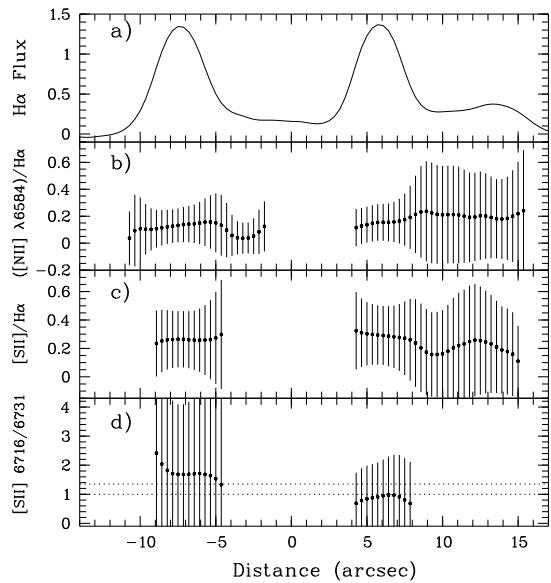}
\caption{H$\alpha$ line intensity
(in 10$^{-15}$ erg s$^{-1}$ cm$^{-2}$ \AA $^{-1}$)
vs. distance along the slit (top panel),
and line intensity ratios for PA=150 with SCORPIO.
North-West is to the right. The influence of the strong fringing
in the SCORPIO data causes the very large error bars in the [\ion{S}{ii}]
line ratio and in the [\ion{S}{ii}]/H$\alpha$ ratio.
\label{fig:line_ratios}}
\end{figure}

 We present in Table~\ref{t:Intens} the line intensities F($\lambda$)
of all relevant emission lines measured in the 1D spectra,
extracted from 2D UAGS spectrum and normalized
by the intensity of H$\beta$ and I($\lambda$), corrected for
the foreground extinction C(H$\beta$), and equivalent widths of underlying
Balmer absorption lines EW(abs).
In this table we also present  the electron temperatures for
emission zones of [\ion{O}{iii}] and  [\ion{O}{ii}]
and ionic and total abundances of oxygen,
derived for the above measured line intensities with the classic
\Te\ method, according to the scheme described in \citet{Kni08}  and using the approximations of Izotov et al. (2006).
Since the principal faint line [\ion{O}{iii}]$\lambda$4363 was
not detected in our data we applied the so-called ``semi-empirical method'',
where the tight correlation between T(\ion{O}{iii}) and the total intensity of
the strong oxygen lines relative to I(H$\beta$) was used \citep{Pag79,Shav83}.

After T(\ion{O}{iii}) was estimated from the total intensity of
[\ion{O}{ii}]$\lambda$3727 and [\ion{O}{iii}]$\lambda\lambda$4959, 5007 lines,
all the other calculations were made as in the classic \Te\ method, using the measured
line intensities.  Since the [\ion{S}{ii}] $\lambda$6717/$\lambda$6731 line ratio is higher than 1.4
for region {\it 2} and is not measurable
for the other areas,  we adopted in all our calculations an electron density N$_e$=10 cm$^{-2}$.
We note that using different N$_e$ values, such as N$_e=100$ cm$^{-2}$, would not have changed
the final results since the differences would be smaller than the errors in the line
intensities.

The areas studied with the 2D UAGS spectrum are shown in Figure~\ref{Fig:2D-spectrum}.
Our calculations yielded A$_B\simeq$2.1--3.8 mag, a rather high value
for disk galaxies but not an exceptional one \citep[see Figure 3 in][]{Zee98}.
Using the inclination-extinction relation from Giovanelli et al. (1994)
for spiral galaxies, transformed to the B-band assuming a Milky Way extinction law,
would have predicted only A$_B\simeq$0.95.

The oxygen abundance values for the different regions, as derived from our data,
do not differ within the rms uncertainties, with weighted mean of
12+log(O/H)$\sim$8.49$\pm$0.08 dex.
Although this is not a low abundance, it is somewhat  lower than would have been
expected using the ring brightness measured here and the metallicity-luminosity
relation of Melbourne \& Salzer (2002) for Kitt Peak National Observatory International Spectroscopy Survey (KISS) galaxies.
We emphasize out that we do not see an abundance gradient with galactocentric distance as seen
 in large spirals, e.g., by \citet{Zee98}, where
all disk galaxies studied show strong radial oxygen abundance gradients.

While the interference fringes in the SCORPIO spectra are strong, and the quality
of the [\ion{S}{ii}]/H$\alpha$ ratio is bad,
the [\ion{N}{ii}]/H$\alpha$ ratio is everywhere better than the [\ion{S}{ii}]/H$\alpha$ ratio,
and it is possible to see from Figure~\ref{fig:line_ratios} that this ratio is very uniform,
indicating the stability of abundance determinations along the radius in RG1.

\subsection{General remarks}

We can estimate the total dynamical mass of RG1 from the gas rotation curve; this reaches an
asymptotic value of $\sim$300 km s$^{-1}$ that is maintained out to $\sim$14.5 kpc. Correcting
the asymptotic velocity for inclination, the total mass within the outermost extent of the ring
is therefore $\sim4 \times 10^{11}$ M$_{\odot}$. Since the rotation curve is $\sim$flat at distances farther
than 15 arcsec=18 kpc while the light intensity is
significantly reduced within the ring, this is another evidence for the presence of significant amounts of dark matter
that influences the dynamics. 

We compared the properties of RG1 with those of the massive disk systems studied by Noordermeer \&
Verheijen (2007), specifically regarding the location on their baryonic Tully-Fisher plot (middle panel in  their Figure 6), and found that RG1 lies slightly above, but not extremely so, from their fitted relation. Note that we adopted v$_{\rm max}$=305 km s$^{-1}$ for this comparison. We also compared with the similar properties of PRGs in the right panels of Figures 1 and 2 of Arnaboldi et al. (2004), adopting $\Delta v_{20}$=610 km s$^{-1}$. RG1 lies on the right dashed line in their Figure 1, near the top of distribution,
 considerably to the right of the spiral galaxies line plotted in
their Figure 2, and on the same line as the S0 galaxies. The CB of RG1 lies also on the same line, using
v$_{\rm max}$=205 km s$^{-1}$ and the total magnitude measured here.

\section{Conclusions}
\label{txt:summ}

We presented above SDSS photometry, long-slit spectroscopy, and FPI data about a possible ring galaxy. The observations have shown that:

\begin{enumerate}
\item The central body shows a S\'{e}rsic surface brightness distribution with $n\simeq1.2$, is much redder than the ring, and exhibits emission lines on top of an absorption spectrum.

\item The ring is blue, with strong emission lines, maintains the same colour through its width,
and is connected to the CB by a linear structure. The H${\alpha}$-to-H${\beta}$ emission line ratio indicates A$_B$=2.7 in the ring. This, and the forbidden line ratios, indicate that the
ring metallicity is 12+log(O/H)$\sim$8.32, lower than expected considering the ring luminosity.

\item The CB stars have a velocity dispersion commensurate with the maximal ring rotation velocity. Their rotation velocity at the outermost locations joins smoothly with the ring rotation.

\item The ionized gas in the ring shows an increasing rotation curve almost to its outermost point. The systemic velocity of each ring segment decreases with galacto-centric distance. The plane of the ring is tilted by 58$^{\circ}\pm10^{\circ}$, or 73$^{\circ}\pm11^{\circ}$, from the plane of the CB.

\item The rotation curve indicates a total dynamical mass of 4$\times 10^{11}$ M$_{\odot}$ within the outer edge of the ring. Since the entire object has M$_g$=--20.35, assuming  a solar absolute magnitude of 5.45 mag in the same spectral band implies a total object luminosity of $\sim2\times 10^{10}$ L$_{\odot}$. Therefore RG1 has an overall M/L$\simeq$20, rather high for galaxies and implying significant amounts of dark matter in the system.
\end{enumerate}

Sil'chenko \& Moiseev (2006) showed that NGC 7742, an Sb galaxy with a nuclear star-forming
ring and lacking a bar, has two global exponential stellar disks with different scale lengths
and a circumnuclear inclined gaseous disk with a radius of 300 pc. The galaxy has a global
gaseous counterrotating subsystem, which they interpret as a result of a past minor merger,
including the appearance of the nuclear star-forming ring lacking a global bar; the ring might
be produced as a resonance feature by tidally-induced oval distortions of the global stellar disk.

The connection between the CB and the ring in RG1, and the smooth match of the rotation curves of the CB (from stellar absorption lines) to that of the ring (from the emission lines), indicates that RG1 could probably be an extreme case of barred spiral where the two arms form a complete circle. However, the kinematics described here, in particular the second solution obtained for the inclination, would indicate an extreme warp of the disk; while the inner regions are reasonably flat, from about 25 arcsec out to the end of the luminous body the disk turns by 60--70$^{\circ}$ within a few arcsec.

It is also possible that this is another case of polar-ring galaxy, like the well-known NGC4650A
with a relatively  small central stellar body and massive self-gravitating
stellar and gaseous polar ring. The polar disk could have a spiral
structure induced by the non-axisymmetric potential of CB; this is similar to that
of a bar in a disk galaxy but without being a real bar, matching the simulations of Theis et al. (2006).

The similarities with respect to NGC 4650A can be summarized as (a) a CB with a near-exponential profile, with a S\'{e}rsic index $n\simeq1$, (b) a bright and presumably massive, self-gravitating ring or disk, and (c) a spiral or quasi-spiral structure in the polar disk/ring.

 We reconsider here the issue of metal abundance in light of the possible nature of the objet. We remind the
reader that the metal abundance measurement shown in Figure~\ref{fig:line_ratios} showed a fairly constant
abundance of 12+log(O/H)=8.30 to 8.65. The measurements covered ring radii 4$\leq$R$\leq$15 arcsec; given that
the 25 mag arcsec$^{-1}$ surface brightness level is reached at R$\simeq$12 arcsec, this implies that our abundance measurements for the ring cover normalized radii 0.4$\leq \frac{R}{R_{25}} \leq$1.25. We compare this with the
metallicity gradients measured by van Zee et al. (1998) for a sample of spiral galaxies, where their figure 12 shows considerable oxygen gradients over this range.

In contrast to the situation regarding spiral galaxies, as shown by van Zee et al. (1998) as well as by Denicolo et al. (2001), very few PRGs have been studied to the determine their metal abundances. A few examples are UGC 5600 (Shalyapina et al. 2002) where no gradient was found, NGC 7468 (Shalyapina et al. 2004) with an inverted gradient where the outer parts are more metal-rich than the inner parts, and AM1934-563 (Brosch et al. 2007) with a normal gradient for the galaxy but a constant metallicity for the ring.

We conclude that this finding for RG1 vs. a fairly steady gradient for more metal-poor regions further from the centre for disk galaxies and a variety of gradients for PRGs, argues in favour interpreting the extended feature in RG1 as indeed a ring, and not part of a disk.

Another argument in favour of considering RG1 a PRG and not a disky spiral is the character of the kinematics of the gas vs. stars in the central region. Not only have we observed counter-rotation of these two components, but the motions there occur on different planes. This is different for spiral galaxies that show counter-rotation; the motions there take place in the same plane and while counter-rotation is seen along the major axis, only a zero velocity gradient is seen along the minor axis (e.g., Sil'chenko et al. 2009).

We summarize the arguments for or against these two possibilities for the nature of RG1 in the Truth Table~\ref{t:truth}.
The option of a PRG has more + signs than the other option thus, on the balance
of probabilities, we accept it over the alternative of a barred spiral with tight spiral arms and a strong disk warp.
Therefore RG1 could be a rare case of a galaxy where the polar ring shows only a small inclination with respect to the equatorial plane of the CB.

\begin{table}
\caption{Two options to explain RG1} \label{t:truth}
\begin{tabular}{ccc} \\ \hline
  Property        & Disk & PRG   \\ \hline
Elongated CB connected with ring  & + & -- \\
Relative brightness ring/CB & -- & +  \\
Colours CB/ring & + & + \\
Ring metal abundance & -- & + \\
Ring radial brightness distribution & -- & + \\
Central gas (H$\beta$) vs. stars kinematics & -- & + \\
CB velocity dispersion & -- & + \\
Tully-Fisher & + & -- \\
Cluster neighbourhood & + & -- \\
           \hline
\end{tabular}
\end{table}



\section*{Acknowledgments}
We acknowledge the discovery of RG1 by Eran Ofek and his willingness to allow us to work on this object. Parts of this paper was written while NB was a sabbatical visitor at the South African Astronomical
Observatory in Cape Town; NB is grateful for this opportunity offered by the SAAO management. We
are grateful for the generous allocation of observing time to this project by the Time Allocation Committee of the Special Astrophysical Observatory.

The 6m SAO RAS telescope is operated under the financial support of  the Ministry of
Science and Education (registration number 01-43). A.M. acknowledges grant no. 09-02-00870 from the Russian Foundation for Basic Research.


Funding for the Sloan Digital Sky Survey (SDSS) has been provided
by the Alfred P. Sloan Foundation, the Participating Institutions,
the National Aeronautics and Space Administration, the National
Science Foundation, the U.S. Department of Energy, the Japanese
Monbukagakusho, and the Max Planck Society. The SDSS Web site is
http://www.sdss.org/.

The SDSS is managed by the Astrophysical Research Consortium (ARC)
for the Participating Institutions. The Participating Institutions
are The University of Chicago, Fermilab, the Institute for
Advanced Study, the Japan Participation Group, The Johns Hopkins
University, Los Alamos National Laboratory, the
Max-Planck-Institute for Astronomy (MPIA), the
Max-Planck-Institute for Astrophysics (MPA), New Mexico State
University, University of Pittsburgh, Princeton University, the
United States Naval Observatory, and the University of Washington.


\bsp

\label{lastpage}

\end{document}